%% file: main.tex
\documentclass{article}

\usepackage{microtype}
\usepackage{graphicx}
\usepackage{booktabs}
\usepackage{amsmath}
\usepackage{amssymb}
\usepackage{xspace}
\usepackage{hyperref}
\usepackage{xurl}
\usepackage[accepted]{mlsys2025}

\usepackage{pifont}
\newcommand{\circled}[1]{\ding{\numexpr171+#1\relax}}

\input{numbers}

\newcommand{\Arch}{Candelta accelerator\xspace}
\newcommand{\Planner}{Candelta planner\xspace}
\newcommand{\E}{\mathbb{E}}
\newcommand{\Normal}{\mathcal{N}}
\newcommand{\tr}{\operatorname{tr}}

\mlsystitlerunning{Manufacturing Differential Sparsity: Planner-Hardware Co-Design for Efficient World Model Planning}

\begin{document}

\twocolumn[
\mlsystitle{Manufacturing Differential Sparsity:\texorpdfstring{\\}{ }Planner-Hardware Co-Design for Efficient\texorpdfstring{\\}{ }World Model Planning}

\begin{mlsysauthorlist}
\mlsysauthor{Wenbo Zhu}{usc}
\mlsysauthor{Paul Yi-Chia Chen}{usc}
\mlsysauthor{Yuanhan Li}{usc}
\mlsysauthor{Zishen Wan}{columbia}
\mlsysauthor{Christopher Torng}{usc}
\end{mlsysauthorlist}

\mlsysaffiliation{usc}{University of Southern California, Los Angeles, CA, USA}
\mlsysaffiliation{columbia}{Columbia University, New York, NY, USA}

\mlsyscorrespondingauthor{Wenbo Zhu}{wenbozhu@usc.edu}

\mlsyskeywords{World Models, Model Predictive Control, Bit-Level Sparsity, Hardware Accelerators, Algorithm-Hardware Co-Design}

\vskip 0.3in

\begin{abstract}
\input{sections/00_abstract}
\end{abstract}
]

\printAffiliationsAndNotice{}

\input{sections/01_intro}
\input{sections/02_background}
\input{sections/03_motivation}
\input{sections/04_planner}
\input{sections/05_hardware}
\input{sections/06_eval}
\input{sections/08_conclusion}

\bibliography{references}
\bibliographystyle{mlsys2025}

\section*{AI Usage Disclosure}

We used Claude as an assistant for coding, figure design and layout, drafting and proofreading suggestions, and reference formatting. The authors designed the methods, wrote the proof structure, specified and verified the baselines, and ran the hardware design flow. The authors reviewed all AI-assisted code and text, decided which suggestions to apply, and take full responsibility for the final work.

\appendix
\input{sections/A_appendix}
\input{sections/A_eval_appendix}

\end{document}

%% file: numbers.tex
\newcommand{\NumP}{300\xspace}
\newcommand{\NumT}{30\xspace}
\newcommand{\NumH}{5\xspace}
\newcommand{\NumE}{30\xspace}
\newcommand{\NumTMACEvent}{1.27\xspace}
\newcommand{\NumCandStepEvals}{45{,}000\xspace}
\newcommand{\NumLeWMParamsM}{18\xspace}
\newcommand{\NumPredShareLeWM}{92.8\%\xspace}
\newcommand{\NumRefStateKiB}{20\,KiB\xspace}
\newcommand{\NumMuOutMiBDINO}{80\,MiB\xspace}
\newcommand{\NumMuOutMiBJEPA}{42\,MiB\xspace}
\newcommand{\NumPERows}{128\xspace}
\newcommand{\NumPECols}{32\xspace}
\newcommand{\NumLanes}{16{,}384\xspace}
\newcommand{\NumFreqGHz}{1\,GHz\xspace}
\newcommand{\NumASRAMMiB}{64\,MiB\xspace}
\newcommand{\NumWTBKiB}{128\,KiB\xspace}
\newcommand{\NumTermCodeBits}{19\xspace}
\newcommand{\NumAccBits}{28\xspace}
\newcommand{\NumDRAMGBs}{25.6\,GB/s\xspace}
\newcommand{\NumRawBitShare}{26\%\xspace}
\newcommand{\NumDenseRows}{208\xspace}
\newcommand{\NumGainCeiling}{2.46$\times$\xspace}
\newcommand{\NumEnergyThorLeWM}{15.6$\times$\xspace}
\newcommand{\NumPredShareDINO}{99.9\%\xspace}
\newcommand{\NumRolloutOverNoneLo}{1.6\xspace}
\newcommand{\NumRolloutOverNoneHi}{2.2$\times$\xspace}
\newcommand{\NumSearchOverMeanMax}{1\%\xspace}
\newcommand{\NumAxisMeanLo}{2.0\xspace}
\newcommand{\NumAxisMeanHi}{2.7$\times$\xspace}
\newcommand{\NumEvBitsNonePusht}{2.03\xspace}
\newcommand{\NumEvBitsMeanPusht}{0.99\xspace}
\newcommand{\NumEvBitsOursPusht}{0.48\xspace}
\newcommand{\NumEvZsMeanPusht}{3.6\%\xspace}
\newcommand{\NumEvZsMeanPushtv}{3.6\xspace}
\newcommand{\NumEvZsOursPusht}{37.8\%\xspace}
\newcommand{\NumEvZsOursPushtv}{37.8\xspace}
\newcommand{\NumEvSuccOursPusht}{88.2\xspace}
\newcommand{\NumEvBitsNoneCube}{2.06\xspace}
\newcommand{\NumEvBitsMeanCube}{1.15\xspace}
\newcommand{\NumEvBitsOursCube}{0.38\xspace}
\newcommand{\NumEvZsMeanCubev}{1.5\xspace}
\newcommand{\NumEvZsOursCubev}{38.2\xspace}
\newcommand{\NumEvSuccOursCube}{64.7\xspace}
\newcommand{\NumEvBitsNoneJEPA}{1.14\xspace}
\newcommand{\NumEvBitsMeanJEPA}{0.48\xspace}
\newcommand{\NumEvBitsOursJEPA}{0.35\xspace}
\newcommand{\NumEvZsMeanJEPAv}{18.4\xspace}
\newcommand{\NumEvZsOursJEPAv}{27.3\xspace}
\newcommand{\NumEvSuccOursJEPA}{61.8\xspace}
\newcommand{\NumEvBitsNoneDINO}{1.62\xspace}
\newcommand{\NumEvBitsMeanDINO}{0.60\xspace}
\newcommand{\NumEvBitsOursDINO}{0.46\xspace}
\newcommand{\NumEvZsMeanDINOv}{12.3\xspace}
\newcommand{\NumEvZsOursDINOv}{19.4\xspace}
\newcommand{\NumEvSuccOursDINO}{63.2\xspace}
\newcommand{\NumEvCutMeanLo}{1.3$\times$\xspace}
\newcommand{\NumEvCutMeanHi}{3.1$\times$\xspace}
\newcommand{\NumEvCutNoneLo}{3.0$\times$\xspace}
\newcommand{\NumEvCutNoneHi}{5.5$\times$\xspace}
\newcommand{\NumEvSpdPrevPusht}{0.91$\times$\xspace}
\newcommand{\NumEvSpdOursCube}{1.62$\times$\xspace}
\newcommand{\NumEvEnPrevJEPA}{0.70\xspace}
\newcommand{\NumEvLimitShareJEPA}{73\%\xspace}
\newcommand{\NumEvSpdPrevDINO}{0.27$\times$\xspace}
\newcommand{\NumEvSpdOursLo}{1.28$\times$\xspace}
\newcommand{\NumEvSpdOursLov}{1.28\xspace}
\newcommand{\NumEvSpdOursHi}{1.80$\times$\xspace}
\newcommand{\NumEvEnOursLo}{0.48\xspace}
\newcommand{\NumEvEnOursHi}{0.62\xspace}
\newcommand{\NumEvSpdMeanLo}{0.81$\times$\xspace}
\newcommand{\NumEvSpdMeanHi}{1.60$\times$\xspace}
\newcommand{\NumEvOursOverMeanLo}{1.12$\times$\xspace}
\newcommand{\NumEvOursOverMeanHi}{1.58$\times$\xspace}
\newcommand{\NumEvOursOverMeanEnLo}{1.08$\times$\xspace}
\newcommand{\NumEvOursOverMeanEnHi}{1.60$\times$\xspace}
\newcommand{\NumEvPrevStateDINOGB}{29\xspace}
\newcommand{\NumEvEqMultLo}{3.8$\times$\xspace}
\newcommand{\NumEvEqMultHi}{5.8$\times$\xspace}
\newcommand{\NumEvBitsNoneTworoom}{2.01\xspace}
\newcommand{\NumEvBitsMeanTworoom}{1.56\xspace}
\newcommand{\NumEvBitsOursTworoom}{0.68\xspace}
\newcommand{\NumEvZsMeanTworoom}{0.5\%\xspace}
\newcommand{\NumEvZsMeanTworoomv}{0.5\xspace}
\newcommand{\NumEvZsOursTworoom}{25.8\%\xspace}
\newcommand{\NumEvZsOursTworoomv}{25.8\xspace}
\newcommand{\NumEvSuccOursTworoom}{86.3\xspace}
\newcommand{\NumEvBitsNoneReacher}{2.12\xspace}
\newcommand{\NumEvBitsMeanReacher}{1.33\xspace}
\newcommand{\NumEvBitsOursReacher}{0.65\xspace}
\newcommand{\NumEvZsMeanReacherv}{1.1\xspace}
\newcommand{\NumEvZsOursReacherv}{24.9\xspace}
\newcommand{\NumEvSuccOursReacher}{73.6\xspace}
\newcommand{\NumEvSpdMeanTworoom}{0.81$\times$\xspace}
\newcommand{\NumEvSpdOursTworoom}{1.28$\times$\xspace}
\newcommand{\NumEvSpdMeanReacher}{0.94$\times$\xspace}
\newcommand{\NumEvSplitGainLo}{4\%\xspace}
\newcommand{\NumEvSplitGainHi}{10\%\xspace}
\newcommand{\NumPraRows}{52\xspace}
\newcommand{\NumEvSpdPraJEPA}{1.29$\times$\xspace}
\newcommand{\NumEvSpdPraDINO}{1.06$\times$\xspace}
\newcommand{\NumEvOursOverPraLo}{1.22$\times$\xspace}
\newcommand{\NumEvOursOverPraLov}{1.22\xspace}
\newcommand{\NumEvOursOverPraHi}{1.58$\times$\xspace}
\newcommand{\NumEvOursOverPraEnLo}{1.2$\times$\xspace}
\newcommand{\NumEvOursOverPraEnHi}{1.6$\times$\xspace}
\newcommand{\NumEvSpdPraLeWMLo}{1.03$\times$\xspace}
\newcommand{\NumEvSpdPraLeWMHi}{1.05$\times$\xspace}
\newcommand{\NumEvEqMultEnLo}{0.34\xspace}
\newcommand{\NumEvEqMultEnHi}{0.45\xspace}
\newcommand{\NumEvSuccDeltaPusht}{+1.5\xspace}
\newcommand{\NumEvSuccCIPusht}{[$-$1.6, +4.7]\xspace}
\newcommand{\NumEvSuccDeltaCube}{+1.5\xspace}
\newcommand{\NumEvSuccCICube}{[$-$1.1, +4.0]\xspace}
\newcommand{\NumEvSuccDeltaTworoom}{+2.0\xspace}
\newcommand{\NumEvSuccCITworoom}{[$-$0.2, +4.4]\xspace}
\newcommand{\NumEvSuccDeltaReacher}{$-$3.5\xspace}
\newcommand{\NumEvSuccCIReacher}{[$-$7.8, +0.8]\xspace}
\newcommand{\NumEvSuccDeltaJEPA}{$-$1.0\xspace}
\newcommand{\NumEvSuccCIJEPA}{[$-$3.4, +1.4]\xspace}
\newcommand{\NumEvSuccDeltaDINO}{$-$1.5\xspace}
\newcommand{\NumEvSuccCIDINO}{[$-$4.0, +1.1]\xspace}
\newcommand{\NumEvQuantDeltaPooled}{+1.1\xspace}
\newcommand{\NumThorLatLeWMv}{147.7\xspace}
\newcommand{\NumThorLatDINOv}{18.8\xspace}
\newcommand{\NumThorLatJEPAv}{28.0\xspace}
\newcommand{\NumThorPowLeWMv}{42\xspace}
\newcommand{\NumThorPowDINOv}{82\xspace}
\newcommand{\NumThorPowJEPAv}{72\xspace}
\newcommand{\NumThorCandLatLeWM}{123\,ms\xspace}
\newcommand{\NumThorCandLatDINO}{43.7\,s\xspace}
\newcommand{\NumThorCandLatJEPA}{25.1\,s\xspace}
\newcommand{\NumThorCandPowLeWM}{3.2\,W\xspace}
\newcommand{\NumThorCandPowDINO}{5.2\,W\xspace}
\newcommand{\NumThorCandPowJEPA}{4.8\,W\xspace}
\newcommand{\NumThorEnGainLeWM}{15.6$\times$\xspace}
\newcommand{\NumThorEnGainDINO}{7.2$\times$\xspace}
\newcommand{\NumThorEnGainJEPA}{17.2$\times$\xspace}
\newcommand{\NumThorSlowDINO}{2.3$\times$\xspace}
\newcommand{\NumEvEnSaveOursLov}{38\xspace}
\newcommand{\NumEvEnSaveOursHi}{52\%\xspace}
\newcommand{\NumEvEnSavePraLov}{16\xspace}
\newcommand{\NumEvEnSavePraHi}{36\%\xspace}
\newcommand{\NumEvPowCubeDensev}{4.15\xspace}
\newcommand{\NumEvPowCubePrav}{3.20\xspace}
\newcommand{\NumEvPowCubeCandv}{3.23\xspace}
\newcommand{\NumLatentCorrLo}{0.81\xspace}
\newcommand{\NumLatentCorrHi}{0.95\xspace}
\newcommand{\NumEvTotAreaDensev}{127.9\xspace}
\newcommand{\NumEvTotAreaPrav}{127.9\xspace}
\newcommand{\NumEvTotAreaCandv}{127.9\xspace}

%% file: sections/00_abstract.tex
JEPA-style latent world models let embodied agents plan toward new goals with model predictive control, but they are still too slow for real-life embodied control: sampling-based MPC repeats predictor inference tens of thousands of times per plan, and even on high-end GPUs one planning step can take tens of seconds. We observe substantial \textbf{latent locality} in this repeated-inference workload: the predictor maps close actions to close latents, and inside a quantized predictor this closeness takes a form that hardware can exploit, \textbf{differential bit-level sparsity}. Unlike an input stream, whose temporal similarity is given, a planner itself decides which candidates it evaluates together, so the planner can choose its own references and, with \textbf{reference forest sampling}, manufacture more sparsity: it couples the sampled action sequences of each group to a shared root, with competitive planning quality. We build the accelerator on the insights of the rich line of bit-level sparsity accelerators and design it and the planner tightly coupled as a single co-design: the planner increases sparsity and exposes structure for hardware scheduling, and the bit-exact hardware effectively uses the sparsity that the planner creates. On three JEPA world models, our co-design is \NumEvSpdOursLov--\NumEvSpdOursHi faster than a dense accelerator and uses \NumEvEnSaveOursLov--\NumEvEnSaveOursHi less energy, and it is \NumEvOursOverPraLov--\NumEvOursOverPraHi faster than prior bit-sparse design and uses \NumEvEnSavePraLov--\NumEvEnSavePraHi less energy, with the same area. To our knowledge, our planner is the first planner that manufactures differential bit-level sparsity across its candidates and exposes its structure for hardware scheduling, and our co-design is the first algorithm-hardware co-design for JEPA-style world-model planning.

%% file: sections/01_intro.tex
\section{Introduction}
\label{sec:intro}

\begin{figure*}[t]
\centering
\includegraphics[width=\textwidth]{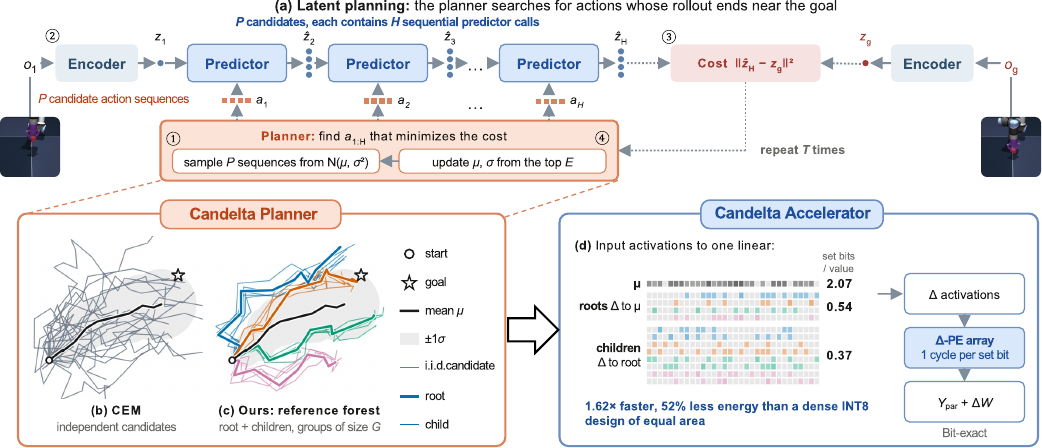}\par
\caption{\textbf{Planner and accelerator design for efficient planning.} (a) A sampling-based planning event: \circled{1}Planner samples P action sequences; \circled{2}$P$ candidates roll out through $H$ sequential predictor calls; \circled{3}Costs are evaluated; \circled{4}The planner refits $\mathcal{N}(\mu, \sigma^2)$ from the $E$ elites. \circled{1}-\circled{4} are repeated for $T$ iterations. (b) Standard CEM rolls out trajectories by sampling the action sequences independently. (c) The \Planner draws from the same sampling distribution as a reference forest: the action sequence of each child is coupled to that of its group's root, while every sampled action sequence keeps the CEM marginal, creating hierarchically correlated candidate trajectories. (d) Each child activation is encoded against its root's, and their roots against the mean candidate.}
\label{fig:teaser}
\end{figure*}

Embodied AI attracts growing attention as learning-based agents become capable of interacting with the physical world. Joint Embedding Predictive Architecture (JEPA) world models are gaining traction as a promising approach to embodied control~\cite{zhou2025dinowm,assran2025vjepa2,maes2026leworldmodel}. While Vision Language Action (VLA) policies~\cite{brohan2023rt2,kim2024openvla,black2024pi0} map observations and instructions directly to actions, JEPA-style latent world models explicitly model how the physical world evolves under given actions. By combining a learned dynamics predictor with strong latent representations and Model Predictive Control (MPC), recent world models achieve online kinematic planning toward new goals without training a task-specific policy, which is closer to deliberate, ``System-2''-style reasoning: the planner can reason about possible futures and select actions based on their predicted outcomes~\cite{lecun2022path,zhou2025dinowm,maes2026leworldmodel}. DINO-WM demonstrates zero-shot planning across several physical environments, while V-JEPA 2-AC extends latent world-model planning to real robotic manipulation~\cite{zhou2025dinowm,assran2025vjepa2}.

Latent world models predict in a lower-dimensional representation space instead of reconstructing high-dimensional signals~\cite{hafner2025dreamerv3}, but they are too slow for real-life embodied control: even on high-end GPUs, one planning step can take tens of seconds~\cite{zhou2025dinowm,assran2025vjepa2}. The reason is that sampling-based MPC repeats inference many times: the planner samples hundreds of action sequences, rolls each one out with the learned predictor, and refines its sampling distribution over many iterations (Figure~\ref{fig:teaser}a). LeWM, for example, makes $300 \times 5 \times 30 = 45{,}000$ predictor calls per plan (300 action sequences, horizon 5, 30 CEM iterations)~\cite{maes2026leworldmodel}. Efficient planning therefore needs efficient execution of this repeated-inference workload.

We first observe substantial redundancy in this computation, rooted in the latent geometry that the world model learns: the predictor maps close actions to close latents, and the latent displacement grows smoothly with the action perturbation (Figure~\ref{fig:latent}). We call this \textbf{latent locality}. Sampling-based planning repeatedly applies action perturbations to trajectories rooted at the same current observation, so latent locality keeps the candidate rollouts close. We show that with a quantized predictor, this closeness can take a form that hardware can exploit, \textbf{differential bit-level sparsity}: when the activation of one predictor invocation is represented as a difference from a related one, a large fraction of its bits are zero.

Prior work has exploited differential bit-level sparsity across temporally adjacent camera inputs in VLA inference~\cite{liu2026deltoris}, where the input stream determines the similarity. In planning, instead, the planner itself decides which action sequences are generated and evaluated, and through latent locality, how close they are. We use this control in two ways. First, the planner supplies a near-optimal reference at no cost: the \emph{mean candidate}, which it evaluates at the mean of its sampling distribution. Input-focused methods cannot identify this reference. Second, the planner can manufacture more sparsity. We introduce \textbf{reference forest sampling}: the planner draws its population as a forest, in which the action sequence of each child is coupled to that of its group's root, while every sampled action sequence keeps the CEM marginal (Figure~\ref{fig:teaser}c). Each child is then encoded against its root, and each root against the mean candidate (Figure~\ref{fig:teaser}d).

Bit-level sparsity has a rich line of accelerator research~\cite{judd2016stripes,albericio2017pragmatic,sharify2019laconic,delmas2019bittactical}. We build the \Arch on their insights, without approximation or compression, and design it and the \Planner as a single method: the hardware uses the structure that the planner exposes for scheduling, and the sparsity that it creates. We call the pair \textbf{Candelta}.

Our contributions are:
\begin{enumerate}
    \item \textbf{Differential sparsity in JEPA world-model planning.}
    We identify and characterize differential bit-level sparsity from latent locality in quantized JEPA-style world-model predictors across candidates, rollout steps and planner refinement iterations, and evaluate its robustness across world models and common tasks, including LeWM, which encodes each observation as one token.

    \item \textbf{Planning-aware references and sparsity-inducing planning.}
    We show that the mean candidate is a free, near-optimal reference for differential sparsity in CEM planning, and we introduce \textbf{reference forest sampling} in the \Planner, which manufactures more sparsity and keeps the CEM marginal and planning quality.

    \item \textbf{Planning-aware differential execution.}
    We co-design the \Planner and the bit-exact \Arch as \textbf{Candelta}, so that the structure the planner exposes becomes the hardware schedule. Candelta is \NumEvSpdOursLov--\NumEvSpdOursHi faster than a dense INT8 accelerator with the same area, at \NumEvEnOursLo--\NumEvEnOursHi of its energy. It is also \NumEvOursOverPraLov--\NumEvOursOverPraHi faster than Pragmatic, the strongest prior bit-sparse design, and uses \NumEvOursOverPraEnLo--\NumEvOursOverPraEnHi less energy. On LeWM, it uses \NumEnergyThorLeWM less energy per planning event than Jetson Thor.
\end{enumerate}
To our knowledge, we are the first to design a JEPA-style world-model planner that manufactures differential bit-level sparsity and exposes its structure for hardware scheduling. This is the first algorithm-hardware co-design for JEPA-style world-model planning.

%% file: sections/02_background.tex
\section{Background and Related Work}
\label{sec:background}

\subsection{Latent World Models}

Recent work has motivated JEPA-style world models as an alternative to reward-driven model-based reinforcement learning or behavioral cloning methods such as VLA models: knowledge of the physical world should come mainly from self-supervised prediction in a regularized latent space~\cite{sobal2025pldm,balestriero2025lejepa,kuang2026lpjepa,murlabadia2026vjepa21,kuang2026lpwm,zhang2026hjepa}. A world model consists of 2 modules: an encoder and a predictor (dynamics model) ~\cite{lecun2022path,assran2023ijepa}. An encoder $\mathrm{enc}_\theta$ maps an observation $o_t$ to a latent, and an action-conditioned predictor $\mathrm{pred}_\phi$ predicts the latent of the next observation,
\begin{equation}
    z_t = \mathrm{enc}_\theta(o_t), \quad \hat{z}_{t+1} = \mathrm{pred}_\phi(z_t, a_t).
\end{equation}
where $a_t$ denotes the action applied at time $t$. The encoder can be a frozen pretrained model, as the DINOv2 patch features of DINO-WM~\cite{oquab2024dinov2,zhou2025dinowm}, or trained end to end with the predictor, as in LeWM, with one token per observation~\cite{maes2026leworldmodel}.

\subsection{Sampling-Based Model Predictive Control}

A planner uses the world model to find the best action sequence: MPC evaluates action sequences with the predictor and replans as the environment evolves. Given an initial observation $o_1$ and a goal observation $o_g$, the encoder first computes the corresponding latent representations
\begin{align}
    \hat{z}_1 = \mathrm{enc}_\theta(o_1), \quad z_g = \mathrm{enc}_\theta(o_g).
\end{align}
For an action sequence $a_{1:H}$ over a planning horizon $H$, the predictor is then rolled out autoregressively according to
\begin{equation}
    \hat{z}_{h+1} = \mathrm{pred}_\phi(\hat{z}_h, a_h).
\end{equation}
The cost is the squared L2 distance between the final predicted latent state and the latent representation of the goal,
\begin{equation}
    C(\hat{z}_H) = \|\hat{z}_H - z_g\|_2^2,
\end{equation}
resulting in the finite-horizon optimization problem
\begin{equation}
    a^*_{1:H} = \arg\min_{a_{1:H}} C(\hat{z}_H).
\end{equation}

Sampling-based optimization, such as Cross-Entropy Method (CEM)~\cite{rubinstein1999cem}, is often preferred for its better performance on navigation and contact-rich manipulation tasks, rather than gradient methods~\cite{zhou2025dinowm,terver2025jepawm}. CEM keeps a Gaussian distribution over action sequences. At CEM iteration $t$, the planner samples $P$ action sequences
\begin{equation}
    \{a^{(i)}_{1:H}\}_{i=1}^{P} \sim \Normal(\mu_{t-1}, \Sigma_{t-1}).
    \label{eq:cem-sample}
\end{equation}
The planner rolls out each sampled action sequence through the world model and computes its planning cost. We call one member of the population a \emph{candidate}: an action sequence together with its rollout (the predicted latent states and the activation vectors of the predictor) and its cost. Candidate $i$ has the action sequence $a^{(i)}_{1:H}$. The $E$ candidates with the lowest costs are selected as elites. The sampling distribution is then updated using statistics of the elites,
\begin{equation}
    \mu_t = \frac{1}{E} \sum_{i \in \mathcal{E}} a^{(i)}_{1:H}, \quad \Sigma_t = \mathrm{Var}_{i \in \mathcal{E}}\left(a^{(i)}_{1:H}\right), \label{eq:cem-mu-sigma}
\end{equation}
where $\mathcal{E}$ denotes the elite set, $|\mathcal{E}| = E$. We call $\mu_t$ the \emph{sampling mean}: an action sequence at the center of the next population. This sampling, rollout, selection, and update process is repeated for $T$ optimization iterations, and the distribution contracts around low-cost action sequences. CEM has many variations such as MPPI~\cite{williams2017mppi} or iCEM~\cite{pinneri2020icem} with different sampling and update rules but the same principles.

This procedure produces a highly structured and compute-intensive inference workload. At every CEM iteration, $P$ candidates are evaluated, and each candidate requires an autoregressive rollout of $H$ predictor steps, across $T$ CEM iterations, resulting in $P \times H \times T$ predictor evaluations. Then, MPC executes the optimized sequence and replans from an updated observation. Recent work makes each predictor evaluation cheaper by reducing the tokens that the predictor processes~\cite{chun2026sparseimagination,kim2026compact}. These methods still compute every candidate independently, and LeWM, with one token per observation, has no tokens for them to remove.

\subsection{Differential Bit-Level Sparsity}
\label{sec:bg-diff}

Consider a linear $y = xW$, where the activation vector $x$ holds $K$ integer values and $W_{k,:}$ holds the weights of input feature $k$. In sign-magnitude form, each value is $x_k = \operatorname{sgn}(x_k) \sum_{j=0}^{B-1} x_{k,j}\, 2^{j}$ with magnitude bits $x_{k,j} \in \{0, 1\}$, so the product reduces to shifts and additions:
\begin{equation}
    y = \sum_{k=1}^{K} x_k\, W_{k,:} = \sum_{k=1}^{K} \sum_{j \in \mathcal{J}(x_k)} \operatorname{sgn}(x_k)\, 2^{j}\, W_{k,:},
    \label{eq:shift-add}
\end{equation}
where $\mathcal{J}(x_k) = \{\, j : x_{k,j} = 1 \,\}$ holds the positions of the set bits. Each summand is a \emph{term}. A bit-parallel multiplier, or a bit-serial accelerator that steps through all $B$ bits~\cite{judd2016stripes}, spends the same work on every value. A term-serial accelerator processes only the terms~\cite{albericio2017pragmatic}, so its work is the number of set bits
\begin{equation}
    b(x) = \sum_{k=1}^{K} |\mathcal{J}(x_k)|,
    \label{eq:set-bits}
\end{equation}
and a zero value costs nothing. Later designs extend this \emph{bit-level sparsity} symmetrically to weights~\cite{delmas2019bittactical,sharify2019laconic,lu2021bitlet,shi2024bitwave,chen2024bbs,guo2025transitive}. In all of them, the operand values alone decide $b(x)$.

Differential computation lets a reference decide $b(x)$ instead. For a linear $y = xW$ and a reference activation vector $x_{\text{ref}}$ whose output $y_{\text{ref}}$ is already known,
\begin{equation}
    y = y_{\text{ref}} + \Delta W, \qquad \Delta = x - x_{\text{ref}},
    \label{eq:bg-delta}
\end{equation}
which is exact in integer arithmetic when $x$ and $x_{\text{ref}}$ share one quantization scale. If $x$ is close to $x_{\text{ref}}$, the values $\delta_k$ of the $\Delta$ vector are small, and a small value has few set bits: $|\delta_k| < 2^m$ gives $|\mathcal{J}(\delta_k)| \le m$. For example, the value $87 = 1010111_2$ has five terms, but against a reference of $85$ its difference $2 = 10_2$ has one. A term-serial accelerator that computes $\Delta W$ therefore does $b(\Delta)$ steps instead of $b(x)$. We call the amount of zero bits of $\Delta$ \emph{differential bit-level sparsity} (Figure~\ref{fig:teaser}(d)).

Sparsity thus depends on the reference. A good $x_{ref}$ is close to $x$, and its $y_{ref}$ is computed first and kept until use. Prior work takes the reference that the input offers: the previous input of a recurrent network or a video stream~\cite{neil2017deltanetworks,cavigelli2017cbinfer}, the previous step of a diffusion model~\cite{kong2024cambricond,kim2025ditto}, or the previous control step of a VLA model~\cite{liu2026deltoris}. In each case, the input decides how close the reference is. In planning, the planner decides the reference, and thus the sparsity (Section~\ref{sec:planner}).

%% file: sections/03_motivation.tex
\section{The Planning Workload}
\label{sec:motivation}

In this section, we first show where the computation of one planning event goes, and then show how the closeness of the candidates gives differential bit-level sparsity.

\subsection{Characterizing Planning Computation}
\label{sec:motivation-time}

Figure~\ref{fig:teaser}(a) shows one planning event. Even LeWM, the most compact model we study (\NumLeWMParamsM\,M parameters, one token per observation~\cite{maes2026leworldmodel}), evaluates $P \times H \times T = \NumCandStepEvals$ candidate steps per event, which is \NumTMACEvent\ TMAC in total, with only 41 MiB of weights. Almost all of the work is in the activation vectors of the candidates. Planning is thus expensive because of the sheer number of dependent predictor calls, rather than the predictor model size.
\begin{figure}[t]
\centering
\includegraphics[width=\columnwidth]{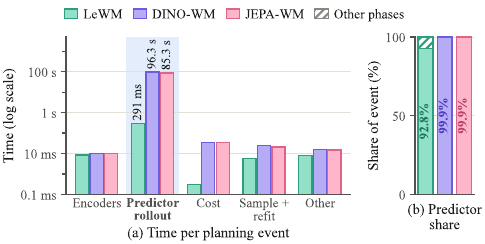}\par
\caption{Profile of the planning computation of three JEPA world models on an NVIDIA A40 GPU. ($P = \NumP$, $H = \NumH$, $T = \NumT$) (a) Time of each phase in one planning event (log scale). (b) Share of the event time that the predictor rollout uses (solid).}
\label{fig:profile}
\end{figure}

Figure~\ref{fig:profile} shows a profile of one planning event on an NVIDIA A40 GPU with optimized kernels and CUDA Graphs. The predictor rollout dominates the planning latency of every model: it uses \NumPredShareDINO of the event for DINO-WM and JEPA-WM, and \NumPredShareLeWM for LeWM. DINO-WM and JEPA-WM are slower than LeWM because each latent state has $16 \times 16$ patch tokens (Appendix~\ref{app:eval-models}). The other phases (encoders, cost, sample + refit) are negligible, so an accelerator must make these predictor calls faster.

\subsection{Differential Sparsity in Planning}
\label{sec:motivation-axes}

\begin{figure}[t]
\centering
\includegraphics[width=\columnwidth]{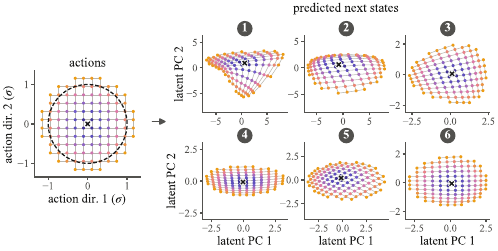}\par
\caption{Left: actions around the sampling mean ($\times$) on a plane through two random directions drawn from the CEM sampling distribution. Right (1--6): 2-D PCA projection of the next-state latents that the predictor computes for these actions from the same current state, for six real LeWM CEM iterations on OGBench Cube~\cite{park2025ogbench}, each on its own scale.}
\label{fig:latent}
\end{figure}

All candidates of a CEM iteration start from the same observation and differ only in their action sequences, which CEM draws as perturbations around the sampling mean at each rollout step. Figure~\ref{fig:latent} shows how the LeWM predictor maps these perturbations one call ahead: the grid of actions becomes a stretched grid of latents that keeps its neighbours, centred on the sampling mean, and the latent displacement grows with the action perturbation (Spearman correlation \NumLatentCorrLo--\NumLatentCorrHi between action distance and latent distance). This \textbf{latent locality} keeps the candidates of one iteration small perturbations of one another inside the predictor, and because the planner chooses the action sequences, it also lets the planner control this closeness (Section~\ref{sec:planner}). We now measure the closeness in the exact form that hardware can exploit: differential bit-level sparsity between the activations of different predictor calls.

\begin{figure}[t]
\centering
\includegraphics[width=\columnwidth]{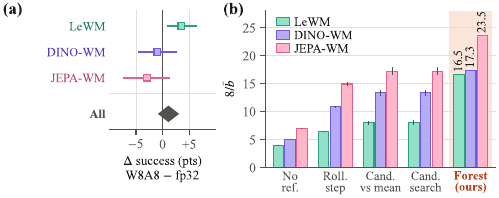}\par
\caption{(a) Impact to planning quality after applying quantization with the 95\% CI. (b) Differential bit-level sparsity when encoded against a reference. Measured on Push-T~\cite{chi2023diffusionpolicy}.}
\label{fig:axes}
\end{figure}

We quantize the predictor weights per channel and activations per tensor with static scales. This quantization does not decrease planning quality, as shown in Figure~\ref{fig:axes}a), with pooled quantized task success \NumEvQuantDeltaPooled points against fp32. We measure $\bar{b}$, the number of set bits $b(\cdot)$ of Eq.~\eqref{eq:set-bits} per value, pooled over all activation vectors, linears and CEM iterations and weighted by the output features that use each value. We report $8/\bar{b}$, the reduction in terms over a bit-serial accelerator that processes all 8 bits. Without a reference, the LeWM activation vectors have \NumRawBitShare of the bits set, which is in line with what earlier work reports for 8-bit DNN activations~\cite{albericio2017pragmatic}. Inherent bit-level sparsity brought by quantization is modest.

Planning offers several axes along which an activation vector has a natural reference (Figure~\ref{fig:axes}b). Along the \emph{rollout-step} axis, activations are encoded against its own previous rollout step predictor call, as temporal differencing does for control inputs~\cite{liu2026deltoris}. This gives only \NumRolloutOverNoneLo to \NumRolloutOverNoneHi fewer set bits than no reference. Along the \emph{candidate} axis, a candidate is encoded against another candidate step-wise. It is the sparser axis for every model: with the \emph{mean candidate}, the candidate conditioned on the mean action sequence as the reference, the activation vectors have \NumAxisMeanLo to \NumAxisMeanHi fewer set bits than with no reference. An online search among the candidate population for a closer reference adds less than \NumSearchOverMeanMax over the mean candidate, so the mean candidate is a near-optimal reference.

%% file: sections/04_planner.tex
\section{\texorpdfstring{Planning-Aware References and Sparsity-Inducing Planning}{Planning-Aware References and Sparsity-Inducing Planning}}
\label{sec:planner}

The candidate axis differs from the references in prior work: Temporal closeness is a property of the input, but the closeness of two candidates is a property of how the planner samples their action sequences. The planner can thus increase differential sparsity without changing the world model.

\subsection{The Mean Anchor and Its Limit}
\label{sec:mean-anchor}

Within iteration $t$, the CEM planner samples action sequences from $\Normal(\mu_t, \Sigma_t)$. The planner also evaluates the mean candidate as candidate~0. We call the encoding of every other candidate against the mean candidate the \textbf{mean anchor}. The mean candidate is near-optimal (Section~\ref{sec:motivation-axes}), because, with $S_t = \tr(\Sigma_t)$,
\begin{equation}
    \E\,\| a^{(i)} - \mu_t \|^2 = S_t,
\end{equation}
while two independently sampled action sequences are $2S_t$ apart. With independent action samples mapped by the predictor into candidates, the mean candidate is thus approximately the closest reference in the population on average. As $S_t$ is controlled by refinements (Eq.\ref{eq:cem-mu-sigma}), no reference does better under CEM. To go further, it must sample action sequences closer to one another than to the sampling mean.

\subsection{Reference Forest Sampling}
\label{sec:forest}

Elite selection and the refit (Eq.~\eqref{eq:cem-mu-sigma}) require that each action sequence has the distribution $\Normal(\mu_t, \Sigma_t)$. The \Planner therefore changes only the joint distribution of the action sequence population, while preserving the distribution of each action sequence.

\begin{figure}[t]
\centering
\includegraphics[width=\columnwidth]{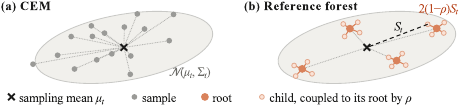}\par
\caption{Comparison between CEM and reference forest at one rollout step, in action space.}
\label{fig:forest}
\end{figure}

\textbf{Structure.} The \Planner arranges the population as a \textbf{reference forest} (Figure~\ref{fig:forest}). Candidate~0 is the mean candidate. The other $P-1$ candidates form groups of $G$. Each group has one \emph{root}, whose action sequence is sampled independently, and up to $G-1$ \emph{children}, whose action sequences are coupled to that of the root.

\textbf{Coupling.} With $a = \mu_t + \sigma_t \odot z$ and $\sigma_t = \operatorname{diag}(\Sigma_t)^{1/2}$, a child uses
\begin{equation}
    z_{\text{child}} = \rho\, z_{\text{root}} + \sqrt{1-\rho^2}\;\epsilon, \qquad \epsilon \sim \Normal(0, I).
    \label{eq:coupling}
\end{equation}
The weights keep the variance of $z_{\text{child}}$ at 1, so each action sequence still follows the distribution $\Normal(\mu_t, \Sigma_t)$ for any $\rho \in [0, 1]$ (Appendix~\ref{app:proof}). The correlation $\rho$ sets how close a child is to its root.

\textbf{Parents.} Each candidate is encoded against its \emph{parent}. The parent of a child is its root. The mean candidate is kept as is. The \Planner decides all parents before the predictor call, so the hardware can compute each parent before its children without data dependency conflicts.

\textbf{Coupling strength.} The expected squared distance from a child to its root is $2(1-\rho)S_t$, and to the sampling mean it is $S_t$ (Figure~\ref{fig:forest}b; Appendix~\ref{app:proof}). The root is thus a closer parent than the mean candidate only when $\rho > 0.5$. Strong coupling is necessary for a gain, but it decreases exploration.

At $\rho = 0$, \Planner exactly degenerates to CEM. Also, because the sampled action sequences remain jointly Gaussian, it can still use the elite set to refit the next iteration sampling distribution just as CEM, and nothing is broken.

\subsection{Adaptive Coupling}
\label{sec:adaptive}

A fixed strong coupling decreases planning quality because it would discourage exploration. With $\rho = 0.8$ in every iteration, the paired median final goal distance is 1.218$\times$ that of standard CEM. We tackle this by introducing adaptive coupling: The \Planner holds the expected squared distance between a child and its root at a budget $\tau$:
\begin{equation}
    \rho_t = \max\left(0,\; 1 - \frac{\tau}{2 S_t}\right).
    \label{eq:rho}
\end{equation}
In early iterations, $S_t$ is large, and $\rho_t$ is near 1. If $\rho_t > \rho_{\max}$, the \Planner makes no groups, and every candidate uses the mean candidate as its parent (the mean anchor). As $S_t$ decreases, $\rho_t$ decreases below $\rho_{\max}$, and coupling starts. Coupling is thus off in the earliest iterations to prevent the lack of exploration resulting in converging to non-optimal solutions (Figure~\ref{fig:ev-coupling}a).

\textbf{Group size.} Larger group size gives fewer roots, which would mean less memory overhead, but more siblings for each root and thus less diverse elites, which are important during exploration, while fewer roots are good for sparsity (Figure~\ref{fig:ev-coupling}b). A \emph{group-size schedule} therefore uses small groups in early iterations and larger groups in late iterations.

\begin{algorithm}[t]
   \caption{Sparsity-inducing CEM with a reference forest}
   \label{alg:planner}
\begin{algorithmic}
   \STATE {\bfseries Input:} $z_1, z_g, \mu_0, \Sigma_0, P, E, G, T, \tau, \rho_{\max}$
   \FOR{$t = 1$ {\bfseries to} $T$}
   \STATE $\rho \leftarrow \max(0, 1 - \tau / 2\tr(\Sigma_{t-1}))$;\; $a^{(0)} \leftarrow \mu_{t-1}$ (mean candidate, sent whole)
   \IF{$\rho \le \rho_{\max}$}
   \FOR{each group: root $r$, children $c$}
   \STATE $z_r \sim \Normal(0, I)$, parent$(r) = 0$
   \STATE $z_c \leftarrow \rho z_r + \sqrt{1-\rho^2}\,\epsilon$, parent$(c) = r$
   \ENDFOR
   \ELSE
   \STATE $z_i \sim \Normal(0, I)$, parent$(i) = 0$ for all $i \ge 1$ \hfill (star)
   \ENDIF
   \STATE $a^{(i)} \leftarrow \mu_{t-1} + \sigma_{t-1} \odot z_i$; roll out all $a^{(i)}$ for $H$ steps,
   \STATE $\mathcal{E} \leftarrow$ $E$ lowest-cost candidates; refit $\mu_t, \Sigma_t$ on $\mathcal{E}$
   \ENDFOR
   \STATE {\bfseries return} $\mu_T$
\end{algorithmic}
\end{algorithm}

%% file: sections/05_hardware.tex
\section{Planning-Aware Differential Execution}
\label{sec:hw}

\begin{figure*}[t]
\centering
\includegraphics[width=\textwidth]{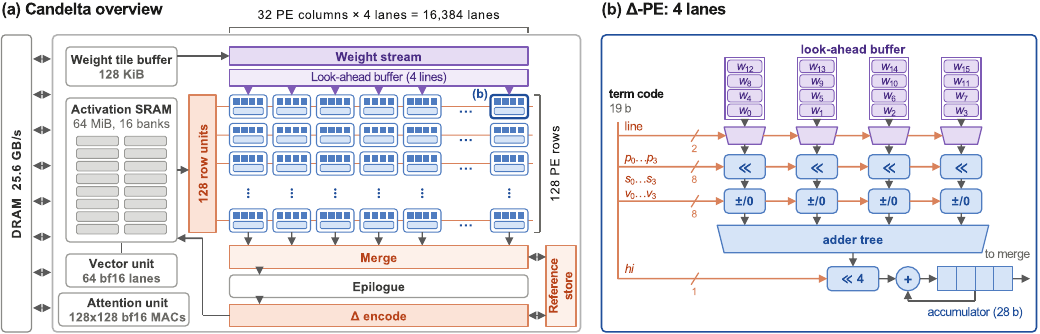}\par
\caption{The \Arch. (a) The chip. Row units send term codes along the PE rows, and the weight stream sends lines down the PE columns. (b) One $\Delta$-PE. The term code from the row unit gives the line, and for each lane $i$ the bit position $p_i \in [0,3]$, the sign $s_i$ and a valid bit $v_i$ of its term. Lane $i$ takes its weight of that line from the look-ahead buffer, shifts it by $p_i$ and applies $s_i$; it adds 0 if $v_i = 0$. If the terms are in bits 4 to 7 (\emph{hi}), one shift by 4 that all lanes share follows the adder tree.}
\label{fig:arch}
\end{figure*}

The \Arch (Figure~\ref{fig:arch}) is a term-serial design for the predictor linears. It builds on earlier term-serial designs ~\cite{albericio2017pragmatic,liu2026deltoris}, but it schedules on the structure that the \Planner produces. It is arithmetically equivalent to dense hardware, and its cycle count follows the sparsity of its operands.

\subsection{Exact Differential Linears}
\label{sec:hw-exact}

Each predictor linear computes $Y = XW$ on $M$ activation vectors of $K$ input features, with $N$ output features. Under the w8a8s scheme, weights are INT8 with one scale per output feature. Activations are INT8 with one static scale per linear input~\cite{jacob2018quantization}, so all $M$ activation vectors of a linear share one scale. Let $q(\cdot)$ be the INT8 code and $p$ the parent of candidate $c$. Then $\Delta = q(x_c) - q(x_p) \in [-254, 254]^K$ is exact, and
\begin{equation}
    Y_c = Y_p + \Delta\, W
    \label{eq:delta}
\end{equation}

is exact in INT32 arithmetic. The \Arch thus gives the same INT8 outputs as the dense integer model.

A difference cannot go through LayerNorm, GELU, softmax or the residual path. The merge unit therefore computes each output with Eq.~\eqref{eq:delta}. The epilogue then dequantizes the output, applies GELU or the residual addition, requantizes it with the static scale of the next linear and subtracts the code of the parent. This step, re-differencing, gives the $\Delta$ vectors of the next linear. LayerNorm and AdaLN run on the vector unit, and attention runs on a dense attention unit, all in bf16 for latent accuracy.

\subsection{The $\Delta$ Array}
\label{sec:hw-core}

The array must spend cycles on set bits, not on values, and share each weight read among many $\Delta$ vectors. Each activation vector occupies one PE row, so we call it a \emph{row} and a $\Delta$ vector a \emph{$\Delta$ row}. The output-stationary array has \NumPERows PE rows, one for each $\Delta$ row, and \NumPECols PE columns, one for each output feature. A \emph{pass} computes one \emph{row tile} (\NumPERows rows) and one \emph{column tile} (\NumPECols output features) over all of $K$.

\textbf{Terms and segments.} The \Arch computes $\Delta W$ term by term (Eq.~\eqref{eq:shift-add}): each set bit of $|\delta|$ adds one weight, shifted and signed. It keeps $\delta$ in 9-bit sign-magnitude form. Each PE has 4 \emph{lanes} (\NumLanes in total, at \NumFreqGHz). The lanes take the 4 values of a \emph{segment}: 4 neighbouring values along $K$. In each cycle, each lane selects the weight of its value, shifts it by 0 to 3 positions and applies the sign. An adder tree adds the lanes into a \NumAccBits-bit accumulator. One shift by 4, which all lanes share, applies to positions 4 to 7, which are rare in $\Delta$. A segment thus takes as many cycles as the largest set-bit count of its values in positions 0--3, plus the largest count in positions 4--7. A segment of zeros takes no cycle.

\textbf{Row units.} Each PE row has one \emph{row unit}. The row unit reads its $\Delta$ row from the activation SRAM as records of 16 values: a 16-bit nonzero mask and the nonzero values. It skips the zero segments and broadcasts one \NumTermCodeBits-bit \emph{term code} per cycle along its PE row. All PEs of a PE row thus do the same terms, and only the PE rows differ in work.

\textbf{Weight stream and look-ahead buffer.} One \emph{weight stream} reads the weight tile buffer in the order of $K$. In each cycle, it reads at most one \emph{line} (the weights of 4 input features for all PE columns) and sends it down the array. All PE rows share a \emph{look-ahead buffer} of 4 lines. A PE row can thus be up to 16 input features ahead of the slowest PE row, and a segment with many set bits stops only the PE rows that need it. The stream stops when the look-ahead buffer is full. Each line is read once per row tile, so a pass takes at least $K/4$ cycles.

\textbf{A Pass.} A pass ends when the slowest PE row finishes. Each PE then sends its sum to the merge after the pass ends. The activation SRAM has 16 banks, and PE row $r$ reads only bank $r \bmod 16$. The epilogue writes each row to its home bank, so no crossbar is necessary.

\subsection{Parent-First Schedule}
\label{sec:hw-schedule}

With the \Arch, the \Planner sets the structure of the rows before the planning event: which rows belong to the mean candidate, to roots and to children, and which parent each child uses. This structure and the shape of the predictor fix the schedule.

\begin{figure}[t]
\centering
\includegraphics[width=\columnwidth]{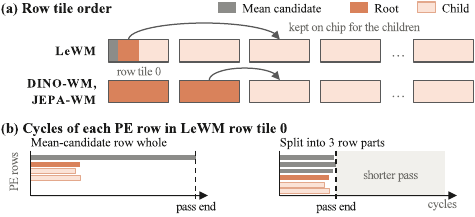}\par
\caption{Parent-first schedule. (a) Row tiles of one linear in execution order: on LeWM, split reference rows share row tile 0 with children; on DINO-WM and JEPA-WM, whole root rows fill the first row tiles. (b) Cycles of each PE row in LeWM row tile 0, with the mean-candidate row whole and split into 3 row parts.}
\label{fig:schedule}
\end{figure}

\textbf{Order.} A $\Delta$ row needs the output of its parent for the merge and the code of its parent for the re-differencing. We call a row of the mean candidate or of a root a \emph{reference row}. In each linear, the reference rows go in the first row tiles and the children after them (Figure~\ref{fig:schedule}a). On LeWM, a reference store keeps outputs and inputs of the reference rows until the merge unit has used them for all children (a store of \NumRefStateKiB). On DINO-WM and JEPA-WM, the activation SRAM keeps them.

\textbf{Balance.} The forest also makes the work of each row predictable. All children of an iteration have the same expected distance to their parents (Section~\ref{sec:forest}), so their set-bit counts are similar. Only the reference rows have many more set bits, and their positions are known. On LeWM, each row of the mean candidate is therefore split into 3 \emph{row parts} and each root row into 2 (Figure~\ref{fig:schedule}b). We don't need to split for DINO-WM and JEPA-WM because their latents are much larger and can fully occupy the hardware.

\textbf{Encoding for each model.} The hardware uses the parent scheme of Section~\ref{sec:forest}. On LeWM, all candidates of a call are on chip, so each root is a $\Delta$ row against the mean candidate. On DINO-WM and JEPA-WM, the \Arch keeps one planner group at a time in the activation SRAM through all transformer blocks. It computes $Q$, $K$ and $V$ one candidate at a time, parent first, and the output projection and the MLP for all members together, so the weights stream once for each transformer block and group. The outputs of the mean candidate for all transformer blocks (\NumMuOutMiBDINO on DINO-WM, \NumMuOutMiBJEPA on JEPA-WM) do not fit on chip next to a group, so the roots go whole and the mean candidate runs as a group of its own.

%% file: sections/06_eval.tex
\section{Evaluation}
\label{sec:eval}

We evaluate how well Candelta turns the closeness of candidate rollouts into speedup and energy savings for planning; each hardware result is for one planning event. After the methodology (Section~\ref{sec:eval-method}), Section~\ref{sec:eval-sparsity} quantifies the sparsity that the \Planner exposes and manufactures, and its effect on task success. Section~\ref{sec:eval-designs} measures the gains of the \Arch against a dense accelerator and prior designs of the same area, and Section~\ref{sec:eval-ablation} isolates each step.

\subsection{Methodology}
\label{sec:eval-method}

\textbf{Models, tasks and planners.} We evaluate LeWM~\cite{maes2026leworldmodel} on Push-T~\cite{chi2023diffusionpolicy}, Cube~\cite{park2025ogbench}, TwoRoom~\cite{sobal2025pldm} and Reacher~\cite{tassa2018dmc}, and JEPA-WM~\cite{terver2025jepawm} and DINO-WM~\cite{zhou2025dinowm} on Push-T (Appendix~\ref{app:eval-models} gives their predictors), all in the w8a8s scheme of Section~\ref{sec:motivation-axes}. As in LeWM, every planning event has $P = \NumP$ candidates, $E = \NumE$ elites, $H = \NumH$ predictor calls per rollout and $T = \NumT$ iterations. We sweep the \Planner configuration on a tuning set, select per model and task the configuration that performs best with the highest sparsity, and report its performance on a held-out test set. The group size, which sets the number of reference rows, is $G = 64$ on LeWM ($G = 32$ on Reacher) and $G = 32$ on DINO-WM. JEPA-WM uses the group-size schedule of Section~\ref{sec:adaptive}: $G = 4$ in iterations 0 to 7 and $G = 32$ after (Appendix~\ref{app:eval-planner}).

\textbf{Hardware model.} A cycle-level simulator of each array replays planner traces at \NumFreqGHz. Energy and area come from post-synthesis of each design in TSMC 28\,nm with Cadence Genus and power analysis with simulated switching activity in Synopsys PrimeTime PX. SRAM energy comes from TSMC memory compiler, and DRAM energy from Micron's LPDDR4 calculator~\cite{micron2018lpddr4power}.

\textbf{Baselines.} All ASICs have the same node, clock and memory system: a \NumASRAMMiB activation SRAM in 16 banks, a \NumWTBKiB weight tile buffer and \NumDRAMGBs of LPDDR4-3200. They also have the same bf16 attention unit, vector unit and epilogue. Only the linear arrays differ (Table~\ref{tab:ev-area}). They have equal area, and each array includes all the hardware it needs. The \emph{dense} array has INT8 MACs, output stationary. The \emph{Pragmatic} array is the strongest raw-operand bit-sparse design for this workload~\cite{albericio2017pragmatic}: its PEs have term-serial lanes, and each PE row moves forward independently. In our simulations, designs that also skip weight bits~\cite{sharify2019laconic,delmas2019bittactical} gain little, because w8a8s weights have little bit-level sparsity. Each chip runs its own best schedule, and all produce arithmetically equivalent results.
To isolate what each design exploits, we also hold the \Arch fixed and vary the policy. A Deltoris-style policy~\cite{liu2026deltoris} subtracts each candidate's activations from the previous predictor call.
As a commercial reference, we measure the same predictor calls on Jetson Thor~\cite{nvidia2025thor}: FP8 in TensorRT, all \NumP candidates of a call in one batch, CUDA graphs and fused attention. We sweep its power modes and subtract its idle power.

\input{tables/ev_area}

\subsection{Planner-Induced Sparsity and Task Performance}
\label{sec:eval-sparsity}

\input{tables/ev_sparsity}

\begin{figure}[t]
\centering
\includegraphics[width=\columnwidth]{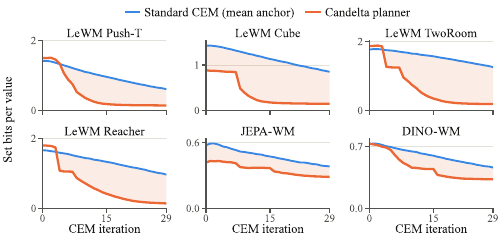}\par
\caption{\textbf{The \Planner removes most set bits once its children are coupled.} Set bits per value in each CEM iteration (MAC-weighted). JEPA-WM and DINO-WM: Push-T.}
\label{fig:ev-sparsity-iter}
\end{figure}

\begin{figure}[t]
\centering
\includegraphics[width=\columnwidth]{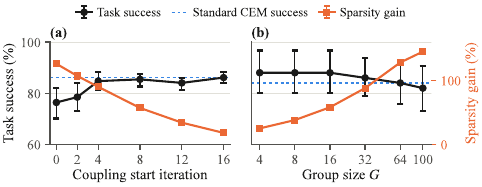}\par
\caption{\textbf{Coupling start iteration and group size.} Task success and sparsity gain over standard CEM on LeWM TwoRoom, against (a) the coupling start iteration ($G = 32$) and (b) the group size $G$.}
\label{fig:ev-coupling}
\end{figure}

The \Planner makes the operands \NumEvCutMeanLo to \NumEvCutMeanHi sparser than standard CEM with the mean anchor, and \NumEvCutNoneLo to \NumEvCutNoneHi sparser than the raw operands (Table~\ref{tab:ev-sparsity}). On LeWM, a value has \NumEvBitsOursPusht, \NumEvBitsOursCube, \NumEvBitsOursTworoom and \NumEvBitsOursReacher set bits on Push-T, Cube, TwoRoom and Reacher, against \NumEvBitsMeanPusht, \NumEvBitsMeanCube, \NumEvBitsMeanTworoom and \NumEvBitsMeanReacher with standard CEM. The \Planner also concentrates the zeros. All-zero 4-value segments, which the \Arch skips at no cost, rise from \NumEvZsMeanPusht to \NumEvZsOursPusht on LeWM Push-T and from \NumEvZsMeanTworoom to \NumEvZsOursTworoom on TwoRoom. Task success stays comparable to standard CEM on every model and task.

Figure~\ref{fig:ev-sparsity-iter} shows when the gain occurs. In the first iterations, the sampling distribution is wide and the children are not coupled yet, so both planners have about the same density. When the coupling starts, the set bits of the \Planner fall quickly. In the late iterations of LeWM Push-T and Cube, a value has about 0.14 set bits. Standard CEM becomes sparser only slowly.

Figure~\ref{fig:ev-coupling} shows how the two main settings of the \Planner trade sparsity gain (over CEM) against task success. Coupling reduces exploration, so it decreases task success when it starts while the sampling distribution still has high variance. Coupling is thus off in early iterations. Group size affects task success less than the coupling start, and $G = 32$ or $64$ balances task success and sparsity. The selected configurations (Appendix~\ref{app:eval-planner}) follow this pattern.

\subsection{Hardware Performance}
\label{sec:eval-designs}

\begin{figure*}[t]
\centering
\includegraphics[width=\textwidth]{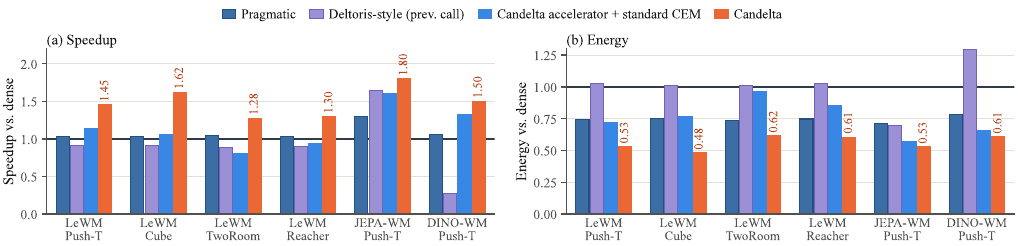}\par
\caption{\textbf{Candelta is the fastest and most energy-efficient design on every model.} Speedup (a) and energy (b) of each design against the dense INT8 accelerator of the same area (line at 1). Pragmatic runs on its own array, and the other policies run on the \Arch. On LeWM TwoRoom and Reacher, the previous-call set bits of the Deltoris-style policy are estimated from those of Push-T and Cube.}
\label{fig:ev-designs}
\end{figure*}

Candelta is \NumEvSpdOursLo to \NumEvSpdOursHi faster than the dense accelerator of the same area on every model and task, LeWM included (\NumEvSpdOursTworoom to \NumEvSpdOursCube), and it uses \NumEvEnOursLo to \NumEvEnOursHi of the dense energy. Prior bit-serial work counts 8 bit-serial lanes as one INT8 MAC and ignores the area overhead of the sparsity hardware~\cite{judd2016stripes,albericio2017pragmatic,chen2024bbs}. Under this equal-multiplier rule, Candelta is \NumEvEqMultLo to \NumEvEqMultHi faster than the dense array, at \NumEvEqMultEnLo to \NumEvEqMultEnHi of its energy (Appendix~\ref{app:eval-eqmult}).

\textbf{Prior bit-serial designs gain little at equal area.} Pragmatic skips the zero bits of the raw operands, but these operands have too few zero bits: \NumEvBitsNoneDINO to \NumEvBitsNoneReacher set bits per value (Table~\ref{tab:ev-sparsity}). Pragmatic is only \NumEvSpdPraLeWMLo--\NumEvSpdPraLeWMHi faster than the dense chip on LeWM and \NumEvSpdPraDINO on DINO-WM. Only on JEPA-WM, which has \NumEvBitsNoneJEPA set bits per value, does it gain more (\NumEvSpdPraJEPA). Candelta is \NumEvOursOverPraLo--\NumEvOursOverPraHi faster than Pragmatic and uses \NumEvOursOverPraEnLo--\NumEvOursOverPraEnHi less energy.

\textbf{Temporal differencing must store state that grows with the population.} The Deltoris-style policy subtracts each candidate's activations in the previous predictor call of its rollout. Its references are therefore the inputs and outputs of every linear, for every candidate: up to \NumEvPrevStateDINOGB\,GB per call on DINO-WM. We count the cost of this state in the same way as for the parents of Candelta, and we give the policy the better of two schedules: the state goes through DRAM, or sets of candidates stay on chip and the weights stream once per set. On LeWM Push-T the state fits on chip, but the latent moves by a full step of the dynamics between calls, so the $\Delta$ rows keep many set bits and the policy reaches only \NumEvSpdPrevPusht of the dense speed. On DINO-WM the state must go through DRAM, and the policy falls to \NumEvSpdPrevDINO. On JEPA-WM the weights stream once per candidate, and the energy rises to \NumEvEnPrevJEPA of the dense chip. In contrast, the references of the \Planner are one mean candidate and a few roots per call, and they stay on chip (Section~\ref{sec:hw-schedule}). A Ditto-style policy, which subtracts the same candidate index in the previous CEM iteration, is impractical: it must store even more state, and CEM draws all candidates again in each iteration, so its references are not related.

\textbf{The \Planner beats standard CEM on the same chip.} With standard CEM and the mean anchor, the \Arch reaches \NumEvSpdMeanLo to \NumEvSpdMeanHi of the dense speed; on LeWM TwoRoom (\NumEvSpdMeanTworoom) and Reacher (\NumEvSpdMeanReacher) it stays slower than the dense chip. The \Planner adds a factor of \NumEvOursOverMeanLo--\NumEvOursOverMeanHi in speedup and \NumEvOursOverMeanEnLo--\NumEvOursOverMeanEnHi in energy. The gain is largest on LeWM, where coupling removes the most set bits (Figure~\ref{fig:ev-sparsity-iter}).

\textbf{The design limit.} A pass cannot end before the weight stream has sent all $K/4$ lines, so at equal area the \Arch is at most \NumGainCeiling faster than the dense chip (Section~\ref{sec:hw-core}). A child row with few set bits already runs near this floor. The rest of the time goes to the whole rows of the mean candidate and the roots, and to the slowest PE row of each pass. JEPA-WM reaches \NumEvLimitShareJEPA of the design limit.

\input{tables/ev_thor}

\textbf{Candelta uses much less energy than a commercial GPU.} Candelta uses \NumThorEnGainDINO to \NumThorEnGainJEPA less energy per planning event than Jetson Thor, at less than one tenth of its power (Table~\ref{tab:ev-thor}) even without scaling for technology. It is also faster than Thor on LeWM and JEPA-WM. On DINO-WM, the wide linears let Thor use its much larger peak throughput, and Thor is \NumThorSlowDINO faster.

\subsection{Ablation}
\label{sec:eval-ablation}

\begin{figure}[t]
\centering
\includegraphics[width=\columnwidth]{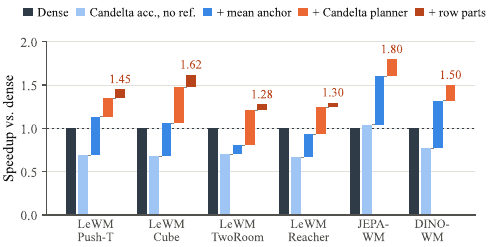}\par
\caption{\textbf{Ablation.} Speedup against the dense chip as each step is added on the \Arch. Row parts: the parent-first balance of Section~\ref{sec:hw-schedule} (LeWM only; the whole reference rows of DINO-WM and JEPA-WM fill their own row tiles).}
\label{fig:ev-waterfall}
\end{figure}

Figure~\ref{fig:ev-waterfall} adds one step at a time. With raw operands, the \Arch is slower than the dense chip on five of six cells. The mean anchor needs no change to the planner and raises every cell, but on LeWM TwoRoom and Reacher it stays below the dense chip. The \Planner then makes every cell faster than the dense chip. On LeWM, the parent-first balance of the reference rows adds a further \NumEvSplitGainLo to \NumEvSplitGainHi. Without it, the whole rows of the mean candidate and the roots set the time of the first row tile of each linear. Each step needs the one before it: the forest is useful only with a reference, and the balance only with a forest whose reference rows are known before the planning event.

%% file: tables/ev_area.tex
\begin{table}[t]
\caption{\textbf{Comparison between the ASICs.} Power: average over one LeWM Cube planning event.}
\label{tab:ev-area}
\vskip 0.1in
\begin{center}
\begin{footnotesize}
\setlength{\tabcolsep}{3pt}
\begin{tabular}{@{}lccc@{}}
\toprule
 & Dense & Pragmatic & Candelta \\
\midrule
PEs (lanes) & $\NumDenseRows{\times}\NumPECols$ (MAC) & $\NumPraRows{\times}\NumPECols$ (16) & $\NumPERows{\times}\NumPECols$ (4) \\
Total area (mm$^2$) & \NumEvTotAreaDensev & \NumEvTotAreaPrav & \NumEvTotAreaCandv \\
Power (W) & \NumEvPowCubeDensev & \NumEvPowCubePrav & \NumEvPowCubeCandv \\
\bottomrule
\end{tabular}
\end{footnotesize}
\end{center}
\vskip -0.1in
\end{table}

%% file: tables/ev_sparsity.tex
\begin{table}[t]
\caption{\textbf{The Candelta planner makes the operands much sparser with competitive task success.} Success: the Candelta planner (Cand.), W8A8, $\Delta$ = Cand.\ $-$ standard CEM. Zero: all-zero 4-value segments. JEPA-WM and DINO-WM: Push-T.}
\label{tab:ev-sparsity}
\vskip 0.1in
\begin{center}
\begin{footnotesize}
\setlength{\tabcolsep}{1.8pt}
\resizebox{\columnwidth}{!}{%
\begin{tabular}{@{}l l rrr rr@{}}
\toprule
 & \multicolumn{1}{c}{Success (\%)} & \multicolumn{3}{c}{Set bits per value} & \multicolumn{2}{c}{Zero (\%)} \\
\cmidrule(lr){2-2} \cmidrule(lr){3-5} \cmidrule(l){6-7}
 & \multicolumn{1}{c}{Cand.\ ($\Delta$, 95\% CI)} & None & Mean & Cand. & Mean & Cand. \\
\midrule
\multicolumn{7}{@{}l}{\textit{LeWM}} \\
\quad Push-T & \NumEvSuccOursPusht(\NumEvSuccDeltaPusht, \NumEvSuccCIPusht) & \NumEvBitsNonePusht & \NumEvBitsMeanPusht & \textbf{\NumEvBitsOursPusht} & \NumEvZsMeanPushtv & \textbf{\NumEvZsOursPushtv} \\
\quad Cube & \NumEvSuccOursCube(\NumEvSuccDeltaCube, \NumEvSuccCICube) & \NumEvBitsNoneCube & \NumEvBitsMeanCube & \textbf{\NumEvBitsOursCube} & \NumEvZsMeanCubev & \textbf{\NumEvZsOursCubev} \\
\quad TwoRoom & \NumEvSuccOursTworoom(\NumEvSuccDeltaTworoom, \NumEvSuccCITworoom) & \NumEvBitsNoneTworoom & \NumEvBitsMeanTworoom & \textbf{\NumEvBitsOursTworoom} & \NumEvZsMeanTworoomv & \textbf{\NumEvZsOursTworoomv} \\
\quad Reacher & \NumEvSuccOursReacher(\NumEvSuccDeltaReacher, \NumEvSuccCIReacher) & \NumEvBitsNoneReacher & \NumEvBitsMeanReacher & \textbf{\NumEvBitsOursReacher} & \NumEvZsMeanReacherv & \textbf{\NumEvZsOursReacherv} \\
JEPA-WM & \NumEvSuccOursJEPA(\NumEvSuccDeltaJEPA, \NumEvSuccCIJEPA) & \NumEvBitsNoneJEPA & \NumEvBitsMeanJEPA & \textbf{\NumEvBitsOursJEPA} & \NumEvZsMeanJEPAv & \textbf{\NumEvZsOursJEPAv} \\
DINO-WM & \NumEvSuccOursDINO(\NumEvSuccDeltaDINO, \NumEvSuccCIDINO) & \NumEvBitsNoneDINO & \NumEvBitsMeanDINO & \textbf{\NumEvBitsOursDINO} & \NumEvZsMeanDINOv & \textbf{\NumEvZsOursDINOv} \\
\bottomrule
\end{tabular}}
\end{footnotesize}
\end{center}
\vskip -0.1in
\end{table}

%% file: tables/ev_thor.tex
\begin{table}[t]
\caption{\textbf{Candelta against Jetson Thor with TensorRT.}}
\label{tab:ev-thor}
\vskip 0.1in
\begin{center}
\begin{footnotesize}
\setlength{\tabcolsep}{4pt}
\begin{tabular}{@{}lccc@{}}
\toprule
 & Latency & Energy & Power \\
 & Thor / Candelta & gain & Thor / Candelta \\
\midrule
LeWM    & \NumThorLatLeWMv{} / \NumThorCandLatLeWM & \NumThorEnGainLeWM & \NumThorPowLeWMv{} / \NumThorCandPowLeWM \\
DINO-WM & \NumThorLatDINOv{} / \NumThorCandLatDINO & \NumThorEnGainDINO & \NumThorPowDINOv{} / \NumThorCandPowDINO \\
JEPA-WM & \NumThorLatJEPAv{} / \NumThorCandLatJEPA & \NumThorEnGainJEPA & \NumThorPowJEPAv{} / \NumThorCandPowJEPA \\
\bottomrule
\end{tabular}
\end{footnotesize}
\end{center}
\vskip -0.1in
\end{table}

%% file: sections/08_conclusion.tex
\section{Conclusion}
\label{sec:conclusion}

In this work, we propose Candelta, a planner-hardware co-design for sampling-based world-model planning. Based on the observation that the predictor maps close actions to close latents, Candelta turns the redundancy of repeated predictor inference into differential bit-level sparsity. The \Planner chooses its own references and, with reference forest sampling, manufactures more sparsity while keeping the CEM marginal and competitive planning quality. The bit-exact \Arch builds on prior bit-level sparsity designs and uses both this sparsity and the structure the planner exposes. On three JEPA world models, Candelta is \NumEvSpdOursLov--\NumEvSpdOursHi faster than a dense INT8 accelerator of the same area, at \NumEvEnOursLo--\NumEvEnOursHi of its energy, and \NumEvOursOverPraLov--\NumEvOursOverPraHi faster than Pragmatic. More broadly, a planner generates its own inference workload and can thus shape it for the hardware. Because the forest changes only how a Gaussian population is drawn, this opportunity extends to other population-based planners such as MPPI and iCEM~\cite{williams2017mppi,pinneri2020icem}, a step toward world-model planning that is fast enough for embodied control.

%% file: sections/A_appendix.tex
\section{Proofs for the Reference Forest}
\label{app:proof}

\textbf{Marginal preservation.} Let $z_{\text{root}} \sim \Normal(0, I)$ and $\epsilon \sim \Normal(0, I)$ be independent, and let $z_{\text{child}} = \rho z_{\text{root}} + \sqrt{1-\rho^2}\,\epsilon$ with $\rho \in [0,1]$. Then $z_{\text{child}} \sim \Normal(0, I)$, and $a_{\text{child}} = \mu_t + \sigma_t \odot z_{\text{child}} \sim \Normal(\mu_t, \Sigma_t)$ for diagonal $\Sigma_t = \operatorname{diag}(\sigma_t^2)$.

\emph{Proof.} $z_{\text{child}}$ is a linear combination of independent Gaussian vectors, so it is Gaussian. Its mean is $\rho \cdot 0 + \sqrt{1-\rho^2}\cdot 0 = 0$. Its covariance is $\rho^2 I + (1-\rho^2) I = I$, because the cross terms vanish by independence. The affine map $z \mapsto \mu_t + \sigma_t \odot z$ then gives mean $\mu_t$ and covariance $\operatorname{diag}(\sigma_t^2) = \Sigma_t$. At $\rho = 0$, $z_{\text{child}} = \epsilon$, so every sampled action sequence is an independent draw, as in standard CEM. $\square$

\textbf{Distances.} For a sampled action sequence $a$ and the sampling mean, $\E\|a - \mu_t\|^2 = \E\|\sigma_t \odot z\|^2 = \sum_d \sigma_{t,d}^2 = S_t$. For two independently sampled action sequences, the difference has covariance $2\Sigma_t$, so the expected squared distance is $2S_t$. For the action sequences of a root and its child, $z_{\text{child}} - z_{\text{root}} = (\rho - 1) z_{\text{root}} + \sqrt{1-\rho^2}\,\epsilon$ has covariance $\left((1-\rho)^2 + 1 - \rho^2\right) I = 2(1-\rho) I$, which gives $\E\|a_{\text{child}} - a_{\text{root}}\|^2 = 2(1-\rho) S_t$. This is smaller than $S_t$ exactly when $\rho > 0.5$. Two children of the same root have $\operatorname{Cov}(z_{c_1}, z_{c_2}) = \rho^2 I$, so the action sequences of siblings are correlated by $\rho^2$, and they are $2(1-\rho^2) S_t$ apart in expectation.

\textbf{Adaptive correlation.} Setting $2(1-\rho) S_t = \tau$ gives $\rho = 1 - \tau / (2 S_t)$; we clip it at 0 and couple with it only when it does not exceed $\rho_{\max} = \sqrt{1 - \eta_{\min}}$. Appendix~\ref{app:eval-planner} gives the selected values.

%% file: sections/A_eval_appendix.tex
\section{Evaluation Details}
\label{app:eval}

\subsection{World-Model Predictors}
\label{app:eval-models}

Table~\ref{tab:predictors} gives the predictors of the three world models. All three are transformers with 6 blocks, and each block has four main linears: QKV, attention output, MLP up and MLP down. LeWM and JEPA-WM condition each block on the action through AdaLN. DINO-WM has no AdaLN: it joins the action and proprioception embeddings to the features of each patch token, so its width is 384 + 10 + 20 = 414. We use the DINO-WM and JEPA-WM Push-T models of the JEPA-WM release~\cite{terver2025jepawm}.

\begin{table}[ht]
\caption{Predictors of the evaluated world models. Latent shape: tokens $\times$ features of one observation at the encoder output. Parameters: the 6 transformer blocks, AdaLN included.}
\label{tab:predictors}
\vskip 0.1in
\begin{center}
\begin{footnotesize}
\setlength{\tabcolsep}{3pt}
\begin{tabular}{@{}l c c c@{}}
\toprule
 & LeWM & DINO-WM & JEPA-WM \\
\midrule
Encoder & ViT-Tiny & DINOv2 ViT-S & DINOv2 ViT-S \\
Latent shape & $1 \times 192$ & $16 \times 16 \times 384$ & $16 \times 16 \times 384$ \\
Action input & AdaLN & per token & AdaLN \\
Blocks $\times$ width & $6 \times 192$ & $6 \times 414$ & $6 \times 400$ \\
Heads $\times$ head size & $16 \times 64$ & $16 \times 64$ & $16 \times 25$ \\
MLP width & 2048 & 2048 & 1600 \\
Parameters & 10.8\,M & 20.4\,M & 17.3\,M \\
\bottomrule
\end{tabular}
\end{footnotesize}
\end{center}
\vskip -0.1in
\end{table}

\subsection{Planner Configurations}
\label{app:eval-planner}

For each model and task, we sweep the \Planner configuration on tuning episodes and keep the configuration that performs best with the highest sparsity. The tuning episodes are not in the test set. Table~\ref{tab:planner-config} lists the selected configurations. All of them use $P = \NumP$, $E = \NumE$, $H = \NumH$ and $T = \NumT$, and every candidate is encoded against its parent (Section~\ref{sec:forest}).

The \Planner sets $\rho_t$ with Eq.~(\ref{eq:rho}) and the budget $\tau$, given as a fraction of $D$, the number of values in one action sequence, and couples only when $\rho_t \le \rho_{\max} = \sqrt{1 - \eta_{\min}}$ (Algorithm~\ref{alg:planner}).

\begin{table}[ht]
\caption{Selected \Planner configurations. $G$: group size (JEPA-WM: $G = 4$ in iterations 0 to 7, then 32). $\tau / D$: distance budget per value of the action sequence.}
\label{tab:planner-config}
\vskip 0.1in
\begin{center}
\begin{footnotesize}
\setlength{\tabcolsep}{2.2pt}
\begin{tabular}{@{}l r r r@{}}
\toprule
Model, task & $G$ & $\tau / D$ & $\rho_{\max}$ \\
\midrule
LeWM Push-T  & 64 & 0.05 & 0.964 \\
LeWM Cube    & 64 & 0.1  & 0.964 \\
LeWM TwoRoom & 64 & 0.1  & 0.964 \\
LeWM Reacher & 32 & 0.1  & 0.964 \\
JEPA-WM      & 4, 32 & 0.05 & 0.975 \\
DINO-WM      & 32 & 0.05 & 0.949 \\
\bottomrule
\end{tabular}
\end{footnotesize}
\end{center}
\vskip -0.1in
\end{table}

\subsection{Equal-Multiplier Comparison}
\label{app:eval-eqmult}

Stripes, Pragmatic and BBS compare designs at an equal multiplier count: 8 bit-serial multipliers count as one INT8 multiplier, and the area overhead of the sparsity hardware is ignored~\cite{judd2016stripes,albericio2017pragmatic,chen2024bbs}. Under this rule, the \NumLanes lanes of the \Arch count as $64 \times 32 = 2{,}048$ INT8 MACs. Table~\ref{tab:ev-eqmult} gives Candelta against this dense array. Candelta is \NumEvEqMultLo to \NumEvEqMultHi faster than the equal-multiplier array. The rule favors the sparse design, so equal area stays our main rule.

\input{tables/ev_eqmult}

%% file: tables/ev_eqmult.tex
\begin{table}[ht]
\caption{\textbf{Equal-multiplier comparison.} Speedup and energy of the \Arch against a dense array with the same multiplier count ($64 \times 32 = 2{,}048$ INT8 MACs; 8 bit-serial lanes count as one MAC). Equal area (Section~\ref{sec:eval-designs}) is the main rule.}
\label{tab:ev-eqmult}
\vskip 0.1in
\begin{center}
\begin{footnotesize}
\setlength{\tabcolsep}{3pt}
\begin{tabular}{@{}l rrr@{}}
\toprule
 & \multicolumn{3}{c}{Equal multipliers: speedup / energy} \\
\cmidrule(l){2-4}
 & No ref. & CEM & Candelta \\
\midrule
LeWM Push-T & 2.05$\times$\,/\,0.76 & 3.39$\times$\,/\,0.50 & 4.33$\times$\,/\,0.37 \\
LeWM Cube & 2.03$\times$\,/\,0.76 & 3.16$\times$\,/\,0.54 & 4.84$\times$\,/\,0.34 \\
LeWM TwoRoom & 2.09$\times$\,/\,0.75 & 2.41$\times$\,/\,0.68 & 3.81$\times$\,/\,0.43 \\
LeWM Reacher & 2.00$\times$\,/\,0.78 & 2.79$\times$\,/\,0.60 & 3.87$\times$\,/\,0.42 \\
JEPA-WM & 3.38$\times$\,/\,0.56 & 5.18$\times$\,/\,0.41 & 5.82$\times$\,/\,0.38 \\
DINO-WM & 2.49$\times$\,/\,0.71 & 4.26$\times$\,/\,0.49 & 4.85$\times$\,/\,0.45 \\
\bottomrule
\end{tabular}
\end{footnotesize}
\end{center}
\vskip -0.1in
\end{table}

%% file: references.bib
@inproceedings{zhou2025dinowm,
  author    = {Zhou, Gaoyue and Pan, Hengkai and LeCun, Yann and Pinto, Lerrel},
  title     = {{DINO}-{WM}: World Models on Pre-trained Visual Features enable Zero-shot Planning},
  booktitle = {Proceedings of the 42nd International Conference on Machine Learning (ICML)},
  series    = {Proceedings of Machine Learning Research},
  volume    = {267},
  pages     = {79115--79135},
  publisher = {PMLR},
  year      = {2025},
  url       = {https://proceedings.mlr.press/v267/zhou25t.html},
  note      = {arXiv:2411.04983}
}

@article{assran2025vjepa2,
  author    = {Assran, Mahmoud and Bardes, Adrien and Fan, David and Garrido, Quentin and Howes, Russell and Komeili, Mojtaba and Muckley, Matthew and Rizvi, Ammar and Roberts, Claire and Sinha, Koustuv and Zholus, Artem and Arnaud, Sergio and Gejji, Abha and Martin, Ada and Hogan, Francois Robert and Dugas, Daniel and Bojanowski, Piotr and Khalidov, Vasil and Labatut, Patrick and Massa, Francisco and Szafraniec, Marc and Krishnakumar, Kapil and Li, Yong and Ma, Xiaodong and Chandar, Sarath and Meier, Franziska and LeCun, Yann and Rabbat, Michael and Ballas, Nicolas},
  title     = {{V-JEPA} 2: Self-Supervised Video Models Enable Understanding, Prediction and Planning},
  journal   = {arXiv preprint arXiv:2506.09985},
  year      = {2025},
  url       = {https://arxiv.org/abs/2506.09985}
}

@article{maes2026leworldmodel,
  author    = {Maes, Lucas and {Le Lidec}, Quentin and Scieur, Damien and LeCun, Yann and Balestriero, Randall},
  title     = {{LeWorldModel}: Stable End-to-End Joint-Embedding Predictive Architecture from Pixels},
  journal   = {arXiv preprint arXiv:2603.19312},
  year      = {2026},
  url       = {https://arxiv.org/abs/2603.19312}
}

@article{terver2025jepawm,
  author    = {Terver, Basile and Yang, Tsung-Yen and Ponce, Jean and Bardes, Adrien and LeCun, Yann},
  title     = {What Drives Success in Physical Planning with Joint-Embedding Predictive World Models?},
  journal   = {Transactions on Machine Learning Research},
  year      = {2026},
  url       = {https://openreview.net/forum?id=cHZn5Gdh8e},
  note      = {arXiv:2512.24497. Code: https://github.com/facebookresearch/jepa-wms}
}

@inproceedings{assran2023ijepa,
  author    = {Assran, Mahmoud and Duval, Quentin and Misra, Ishan and Bojanowski, Piotr and Vincent, Pascal and Rabbat, Michael and LeCun, Yann and Ballas, Nicolas},
  title     = {Self-Supervised Learning from Images with a Joint-Embedding Predictive Architecture},
  booktitle = {Proceedings of the IEEE/CVF Conference on Computer Vision and Pattern Recognition (CVPR)},
  pages     = {15619--15629},
  year      = {2023},
  doi       = {10.1109/CVPR52729.2023.01499}
}

@misc{lecun2022path,
  author    = {LeCun, Yann},
  title     = {A Path Towards Autonomous Machine Intelligence, Version 0.9.2},
  year      = {2022},
  month     = jun,
  howpublished = {OpenReview},
  url       = {https://openreview.net/forum?id=BZ5a1r-kVsf}
}

@inproceedings{sobal2025pldm,
  author    = {Sobal, Uladzislau and Zhang, Wancong and Cho, Kyunghyun and Balestriero, Randall and Rudner, Tim G. J. and LeCun, Yann},
  title     = {Learning from Reward-Free Offline Data: A Case for Planning with Latent Dynamics Models},
  booktitle = {Advances in Neural Information Processing Systems (NeurIPS)},
  volume    = {38},
  pages     = {43905--43941},
  year      = {2025},
  doi       = {10.52202/085713-1465},
  url       = {https://arxiv.org/abs/2502.14819},
  note      = {arXiv:2502.14819}
}

@article{balestriero2025lejepa,
  author    = {Balestriero, Randall and LeCun, Yann},
  title     = {{LeJEPA}: Provable and Scalable Self-Supervised Learning Without the Heuristics},
  journal   = {arXiv preprint arXiv:2511.08544},
  year      = {2025},
  url       = {https://arxiv.org/abs/2511.08544}
}

@inproceedings{kuang2026lpjepa,
  author    = {Kuang, Yilun and Dagade, Yash and Rudner, Tim G. J. and Balestriero, Randall and LeCun, Yann},
  title     = {Rectified {LpJEPA}: Joint-Embedding Predictive Architectures with Sparse and Maximum-Entropy Representations},
  booktitle = {Proceedings of the 43rd International Conference on Machine Learning (ICML)},
  series    = {Proceedings of Machine Learning Research},
  volume    = {306},
  pages     = {61267--61316},
  publisher = {PMLR},
  year      = {2026},
  url       = {https://proceedings.mlr.press/v306/kuang26b.html},
  note      = {arXiv:2602.01456}
}

@article{kuang2026lpwm,
  author    = {Kuang, Yilun and Dagade, Yash and {Le Lidec}, Quentin and Maes, Lucas and Balestriero, Randall and LeCun, Yann},
  title     = {{LpWM}: A Case for Sparse Representations in World Models},
  journal   = {arXiv preprint arXiv:2608.22764},
  year      = {2026},
  url       = {https://arxiv.org/abs/2608.22764}
}

@article{zhang2026hjepa,
  author    = {Zhang, Wancong and Terver, Basile and Rabbat, Michael and LeCun, Yann and Balestriero, Randall},
  title     = {{H-JEPA}: End-to-End Learning of Hierarchical World Models for Visual Planning},
  journal   = {arXiv preprint arXiv:2610.06805},
  year      = {2026},
  url       = {https://arxiv.org/abs/2610.06805}
}

@inproceedings{murlabadia2026vjepa21,
  author    = {Mur-Labadia, Lorenzo and Muckley, Matthew and Bar, Amir and Assran, Mido and Sinha, Koustuv and Rabbat, Mike and LeCun, Yann and Ballas, Nicolas},
  title     = {{V-JEPA} 2.1: Unlocking Dense Features in Video Self-Supervised Learning},
  booktitle = {Computer Vision -- ECCV 2026},
  series    = {Lecture Notes in Computer Science},
  pages     = {671--689},
  publisher = {Springer Nature Switzerland},
  year      = {2026},
  doi       = {10.1007/978-3-032-37369-4_36}
}

@article{hafner2025dreamerv3,
  author    = {Hafner, Danijar and Pasukonis, Jurgis and Ba, Jimmy and Lillicrap, Timothy},
  title     = {Mastering Diverse Control Tasks through World Models},
  journal   = {Nature},
  volume    = {640},
  number    = {8059},
  pages     = {647--653},
  year      = {2025},
  doi       = {10.1038/s41586-025-08744-2}
}

@article{rubinstein1999cem,
  author    = {Rubinstein, Reuven Y.},
  title     = {The Cross-Entropy Method for Combinatorial and Continuous Optimization},
  journal   = {Methodology and Computing in Applied Probability},
  volume    = {1},
  number    = {2},
  pages     = {127--190},
  year      = {1999},
  doi       = {10.1023/A:1010091220143}
}

@inproceedings{pinneri2020icem,
  author    = {Pinneri, Cristina and Sawant, Shambhuraj and Blaes, Sebastian and Achterhold, Jan and Stueckler, Joerg and Rolinek, Michal and Martius, Georg},
  title     = {Sample-efficient Cross-Entropy Method for Real-time Planning},
  booktitle = {Proceedings of the 2020 Conference on Robot Learning (CoRL)},
  series    = {Proceedings of Machine Learning Research},
  volume    = {155},
  pages     = {1049--1065},
  publisher = {PMLR},
  year      = {2020},
  url       = {https://proceedings.mlr.press/v155/pinneri21a.html}
}

@inproceedings{williams2017mppi,
  author    = {Williams, Grady and Wagener, Nolan and Goldfain, Brian and Drews, Paul and Rehg, James M. and Boots, Byron and Theodorou, Evangelos A.},
  title     = {Information Theoretic {MPC} for Model-Based Reinforcement Learning},
  booktitle = {2017 IEEE International Conference on Robotics and Automation (ICRA)},
  pages     = {1714--1721},
  year      = {2017},
  doi       = {10.1109/ICRA.2017.7989202}
}

@article{oquab2024dinov2,
  author    = {Oquab, Maxime and Darcet, Timoth{\'e}e and Moutakanni, Th{\'e}o and Vo, Huy V. and Szafraniec, Marc and Khalidov, Vasil and Fernandez, Pierre and Haziza, Daniel and Massa, Francisco and El-Nouby, Alaaeldin and Assran, Mido and Ballas, Nicolas and Galuba, Wojciech and Howes, Russell and Huang, Po-Yao and Li, Shang-Wen and Misra, Ishan and Rabbat, Michael and Sharma, Vasu and Synnaeve, Gabriel and Xu, Hu and Jegou, Herv{\'e} and Mairal, Julien and Labatut, Patrick and Joulin, Armand and Bojanowski, Piotr},
  title     = {{DINOv2}: Learning Robust Visual Features without Supervision},
  journal   = {Transactions on Machine Learning Research},
  year      = {2024},
  url       = {https://openreview.net/forum?id=a68SUt6zFt},
  note      = {arXiv:2304.07193}
}

@inproceedings{kim2024openvla,
  author    = {Kim, Moo Jin and Pertsch, Karl and Karamcheti, Siddharth and Xiao, Ted and Balakrishna, Ashwin and Nair, Suraj and Rafailov, Rafael and Foster, Ethan P. and Sanketi, Pannag R. and Vuong, Quan and Kollar, Thomas and Burchfiel, Benjamin and Tedrake, Russ and Sadigh, Dorsa and Levine, Sergey and Liang, Percy and Finn, Chelsea},
  title     = {{OpenVLA}: An Open-Source Vision-Language-Action Model},
  booktitle = {Proceedings of The 8th Conference on Robot Learning (CoRL)},
  series    = {Proceedings of Machine Learning Research},
  volume    = {270},
  pages     = {2679--2713},
  publisher = {PMLR},
  year      = {2024},
  url       = {https://proceedings.mlr.press/v270/kim25c.html}
}

@inproceedings{brohan2023rt2,
  author    = {Zitkovich, Brianna and Yu, Tianhe and Xu, Sichun and Xu, Peng and Xiao, Ted and Xia, Fei and Wu, Jialin and Wohlhart, Paul and Welker, Stefan and Wahid, Ayzaan and Vuong, Quan and Vanhoucke, Vincent and Tran, Huong and Soricut, Radu and Singh, Anikait and Singh, Jaspiar and Sermanet, Pierre and Sanketi, Pannag R. and Salazar, Grecia and Ryoo, Michael S. and Reymann, Krista and Rao, Kanishka and Pertsch, Karl and Mordatch, Igor and Michalewski, Henryk and Lu, Yao and Levine, Sergey and Lee, Lisa and Lee, Tsang-Wei Edward and Leal, Isabel and Kuang, Yuheng and Kalashnikov, Dmitry and Julian, Ryan and Joshi, Nikhil J. and Irpan, Alex and Ichter, Brian and Hsu, Jasmine and Herzog, Alexander and Hausman, Karol and Gopalakrishnan, Keerthana and Fu, Chuyuan and Florence, Pete and Finn, Chelsea and Dubey, Kumar Avinava and Driess, Danny and Ding, Tianli and Choromanski, Krzysztof Marcin and Chen, Xi and Chebotar, Yevgen and Carbajal, Justice and Brown, Noah and Brohan, Anthony and Arenas, Montserrat Gonzalez and Han, Kehang},
  title     = {{RT-2}: Vision-Language-Action Models Transfer Web Knowledge to Robotic Control},
  booktitle = {Proceedings of The 7th Conference on Robot Learning (CoRL)},
  series    = {Proceedings of Machine Learning Research},
  volume    = {229},
  pages     = {2165--2183},
  publisher = {PMLR},
  year      = {2023},
  url       = {https://proceedings.mlr.press/v229/zitkovich23a.html}
}

@article{black2024pi0,
  author    = {Black, Kevin and Brown, Noah and Driess, Danny and Esmail, Adnan and Equi, Michael and Finn, Chelsea and Fusai, Niccolo and Groom, Lachy and Hausman, Karol and Ichter, Brian and Jakubczak, Szymon and Jones, Tim and Ke, Liyiming and Levine, Sergey and Li-Bell, Adrian and Mothukuri, Mohith and Nair, Suraj and Pertsch, Karl and Shi, Lucy Xiaoyang and Tanner, James and Vuong, Quan and Walling, Anna and Wang, Haohuan and Zhilinsky, Ury},
  title     = {{$\pi_0$}: A Vision-Language-Action Flow Model for General Robot Control},
  journal   = {arXiv preprint arXiv:2410.24164},
  year      = {2024},
  url       = {https://arxiv.org/abs/2410.24164}
}

@inproceedings{judd2016stripes,
  author    = {Judd, Patrick and Albericio, Jorge and Hetherington, Tayler and Aamodt, Tor M. and Moshovos, Andreas},
  title     = {Stripes: Bit-serial Deep Neural Network Computing},
  booktitle = {2016 49th Annual IEEE/ACM International Symposium on Microarchitecture (MICRO)},
  pages     = {1--12},
  year      = {2016},
  doi       = {10.1109/MICRO.2016.7783722}
}

@inproceedings{albericio2017pragmatic,
  author    = {Albericio, Jorge and Delm{\'a}s, Alberto and Judd, Patrick and Sharify, Sayeh and O'Leary, Gerard and Genov, Roman and Moshovos, Andreas},
  title     = {Bit-Pragmatic Deep Neural Network Computing},
  booktitle = {Proceedings of the 50th Annual IEEE/ACM International Symposium on Microarchitecture (MICRO)},
  pages     = {382--394},
  year      = {2017},
  doi       = {10.1145/3123939.3123982}
}

@inproceedings{delmas2019bittactical,
  author    = {Delmas Lascorz, Alberto and Judd, Patrick and Malone Stuart, Dylan and Poulos, Zissis and Mahmoud, Mostafa and Sharify, Sayeh and Nikolic, Milos and Siu, Kevin and Moshovos, Andreas},
  title     = {Bit-Tactical: A Software/Hardware Approach to Exploiting Value and Bit Sparsity in Neural Networks},
  booktitle = {Proceedings of the 24th International Conference on Architectural Support for Programming Languages and Operating Systems (ASPLOS)},
  pages     = {749--763},
  year      = {2019},
  doi       = {10.1145/3297858.3304041}
}

@inproceedings{sharify2019laconic,
  author    = {Sharify, Sayeh and Delmas Lascorz, Alberto and Mahmoud, Mostafa and Nikolic, Milos and Siu, Kevin and Malone Stuart, Dylan and Poulos, Zissis and Moshovos, Andreas},
  title     = {Laconic Deep Learning Inference Acceleration},
  booktitle = {Proceedings of the 46th International Symposium on Computer Architecture (ISCA)},
  pages     = {304--317},
  year      = {2019},
  doi       = {10.1145/3307650.3322255}
}

@inproceedings{lu2021bitlet,
  author    = {Lu, Hang and Chang, Liang and Li, Chenglong and Zhu, Zixuan and Lu, Shengjian and Liu, Yanhuan and Zhang, Mingzhe},
  title     = {Distilling Bit-level Sparsity Parallelism for General Purpose Deep Learning Acceleration},
  booktitle = {MICRO-54: 54th Annual IEEE/ACM International Symposium on Microarchitecture},
  pages     = {963--976},
  year      = {2021},
  doi       = {10.1145/3466752.3480123}
}

@inproceedings{shi2024bitwave,
  author    = {Shi, Man and Jain, Vikram and Joseph, Antony and Meijer, Maurice and Verhelst, Marian},
  title     = {{BitWave}: Exploiting Column-Based Bit-Level Sparsity for Deep Learning Acceleration},
  booktitle = {2024 IEEE International Symposium on High-Performance Computer Architecture (HPCA)},
  pages     = {732--746},
  year      = {2024},
  doi       = {10.1109/HPCA57654.2024.00062}
}

@inproceedings{chen2024bbs,
  author    = {Chen, Yuzong and Meng, Jian and Seo, Jae-sun and Abdelfattah, Mohamed S.},
  title     = {{BBS}: Bi-Directional Bit-Level Sparsity for Deep Learning Acceleration},
  booktitle = {2024 57th IEEE/ACM International Symposium on Microarchitecture (MICRO)},
  pages     = {551--564},
  year      = {2024},
  doi       = {10.1109/MICRO61859.2024.00048}
}

@inproceedings{kong2024cambricond,
  author    = {Kong, Weihao and Hao, Yifan and Guo, Qi and Zhao, Yongwei and Song, Xinkai and Li, Xiaqing and Zou, Mo and Du, Zidong and Zhang, Rui and Liu, Chang and Wen, Yuanbo and Jin, Pengwei and Hu, Xing and Li, Wei and Xu, Zhiwei and Chen, Tianshi},
  title     = {{Cambricon-D}: Full-Network Differential Acceleration for Diffusion Models},
  booktitle = {2024 ACM/IEEE 51st Annual International Symposium on Computer Architecture (ISCA)},
  pages     = {903--914},
  year      = {2024},
  doi       = {10.1109/ISCA59077.2024.00070}
}

@inproceedings{kim2025ditto,
  author    = {Kim, Sungbin and Lee, Hyunwuk and Cho, Wonho and Park, Mincheol and Ro, Won Woo},
  title     = {Ditto: Accelerating Diffusion Model via Temporal Value Similarity},
  booktitle = {2025 IEEE International Symposium on High Performance Computer Architecture (HPCA)},
  pages     = {338--352},
  year      = {2025},
  doi       = {10.1109/HPCA61900.2025.00035}
}

@article{liu2026deltoris,
  author    = {Liu, Zheng and Guo, Zeyu and Liu, Zihan and Wu, Anbang and Zhao, Han and Liu, Fangxin and He, Zhezhi and Han, Yinhe and Leng, Jingwen and Guo, Minyi and Gan, Yiming and Feng, Yu},
  title     = {Deltoris: Enabling Real-time {VLA} Inference in Embodied {AI} via Bit-level Sparsity and Speculative Inference},
  journal   = {arXiv preprint arXiv:2608.04428},
  year      = {2026},
  url       = {https://arxiv.org/abs/2608.04428}
}

@inproceedings{guo2025transitive,
  author    = {Guo, Cong and Wei, Chiyue and Tang, Jiaming and Duan, Bowen and Han, Song and Li, Hai and Chen, Yiran},
  title     = {Transitive Array: An Efficient {GEMM} Accelerator with Result Reuse},
  booktitle = {Proceedings of the 52nd Annual International Symposium on Computer Architecture (ISCA)},
  pages     = {990--1004},
  year      = {2025},
  doi       = {10.1145/3695053.3731043}
}

@inproceedings{neil2017deltanetworks,
  author    = {Neil, Daniel and Lee, Jun Haeng and Delbruck, Tobi and Liu, Shih-Chii},
  title     = {Delta Networks for Optimized Recurrent Network Computation},
  booktitle = {Proceedings of the 34th International Conference on Machine Learning (ICML)},
  series    = {Proceedings of Machine Learning Research},
  volume    = {70},
  pages     = {2584--2593},
  publisher = {PMLR},
  year      = {2017},
  url       = {https://proceedings.mlr.press/v70/neil17a.html}
}

@inproceedings{cavigelli2017cbinfer,
  author    = {Cavigelli, Lukas and Degen, Philippe and Benini, Luca},
  title     = {{CBinfer}: Change-Based Inference for Convolutional Neural Networks on Video Data},
  booktitle = {Proceedings of the 11th International Conference on Distributed Smart Cameras (ICDSC)},
  pages     = {1--8},
  year      = {2017},
  doi       = {10.1145/3131885.3131906},
  note      = {arXiv:1704.04313}
}

@misc{nvidia2025thor,
  author    = {{NVIDIA Corporation}},
  title     = {{NVIDIA Jetson AGX Thor} and {Jetson Thor} Series Modules},
  year      = {2025},
  howpublished = {\url{https://www.nvidia.com/en-us/autonomous-machines/embedded-systems/jetson-thor/}},
  note      = {Accessed 2026-10-14}
}

@inproceedings{jacob2018quantization,
  author    = {Jacob, Benoit and Kligys, Skirmantas and Chen, Bo and Zhu, Menglong and Tang, Matthew and Howard, Andrew and Adam, Hartwig and Kalenichenko, Dmitry},
  title     = {Quantization and Training of Neural Networks for Efficient Integer-Arithmetic-Only Inference},
  booktitle = {Proceedings of the IEEE Conference on Computer Vision and Pattern Recognition (CVPR)},
  pages     = {2704--2713},
  year      = {2018},
  doi       = {10.1109/CVPR.2018.00286}
}

@inproceedings{chun2026sparseimagination,
  author    = {Chun, Junha and Jeong, Youngjoon and Kim, Taesup},
  title     = {Sparse Imagination for Efficient Visual World Model Planning},
  booktitle = {International Conference on Learning Representations (ICLR)},
  year      = {2026},
  url       = {https://arxiv.org/abs/2506.01392},
  note      = {arXiv:2506.01392}
}

@inproceedings{kim2026compact,
  author    = {Kim, Dongwon and Seo, Gawon and Lee, Jinsung and Cho, Minsu and Kwak, Suha},
  title     = {Planning in 8 Tokens: A Compact Discrete Tokenizer for Latent World Model},
  booktitle = {Proceedings of the IEEE/CVF Conference on Computer Vision and Pattern Recognition (CVPR)},
  year      = {2026},
  pages     = {8183--8193},
  url       = {https://arxiv.org/abs/2603.05438},
  note      = {arXiv:2603.05438}
}

@misc{micron2018lpddr4power,
  author       = {{Micron Technology}},
  title        = {{TN-53-01}: {LPDDR4} System Power Calculator},
  year         = {2018},
  howpublished = {Spreadsheet, rev.~2.1}
}

@inproceedings{chi2023diffusionpolicy,
  author    = {Chi, Cheng and Feng, Siyuan and Du, Yilun and Xu, Zhenjia and Cousineau, Eric and Burchfiel, Benjamin and Song, Shuran},
  title     = {Diffusion Policy: Visuomotor Policy Learning via Action Diffusion},
  booktitle = {Proceedings of Robotics: Science and Systems (RSS)},
  year      = {2023},
  doi       = {10.15607/RSS.2023.XIX.026}
}

@inproceedings{park2025ogbench,
  author    = {Park, Seohong and Frans, Kevin and Eysenbach, Benjamin and Levine, Sergey},
  title     = {{OGBench}: Benchmarking Offline Goal-Conditioned {RL}},
  booktitle = {International Conference on Learning Representations (ICLR)},
  year      = {2025},
  url       = {https://arxiv.org/abs/2410.20092},
  note      = {arXiv:2410.20092}
}

@article{tassa2018dmc,
  author    = {Tassa, Yuval and Doron, Yotam and Muldal, Alistair and Erez, Tom and Li, Yazhe and {de Las Casas}, Diego and Budden, David and Abdolmaleki, Abbas and Merel, Josh and Lefrancq, Andrew and Lillicrap, Timothy and Riedmiller, Martin},
  title     = {{DeepMind} Control Suite},
  journal   = {arXiv preprint arXiv:1801.00690},
  year      = {2018},
  url       = {https://arxiv.org/abs/1801.00690}
}
